\documentclass[12pt, oneside]{book}

\usepackage[a4paper, left=35mm,top=25mm,bottom=25mm]{geometry}
\usepackage[pdftex]{graphicx}
\usepackage{amsmath}
\usepackage{amssymb}
\usepackage{float}
\usepackage{tipa}
\usepackage{multirow}
\usepackage{textcomp}
\usepackage{amsthm}

\usepackage{subcaption}
\usepackage[table,xcdraw]{xcolor}
\usepackage{setspace}
\DeclareUnicodeCharacter{202F}{\,}
\usepackage{appendix}
\begin{document}

\pagestyle{plain}
\frontmatter
%



%
\mainmatter

\begin{titlepage}
\enlargethispage{3cm}

\begin{center}

\vspace*{-2cm}

\textbf{\large EFFECT OF COVID-19 LOCKDOWN \\ \vspace{-0.5pt} ON HUMAN BEHAVIOUR \\ \vspace{3.5pt} USING  ANALYTICAL HIERARCHY PROCESS}\\
\vspace{1.5 cm}
 A PROJECT REPORT \\
 SUBMITTED IN PARTIAL FULFILLMENT OF THE REQUIREMENTS  \\
  FOR THE AWARD OF THE DEGREE \\
  OF\\
  MASTER OF SCIENCE \\
  IN\\[10pt]

 {\Large \bf APPLIED MATHEMATICS}\\[5pt]
 \vspace{1.5 cm}
{\normalsize {Submitted By:}}\\[5pt]
{\large\bf {Mansi Yadav}}\\[5pt]
{\large (2K21/MSCMAT/30)}\\
{\large\bf {Rashi Jain}}\\[5pt]
{\large (2K21/MSCMAT/40)}\\ [10pt]
{\normalsize {Under the supervision of}}\\[5pt]
{\large\bf {PROF. ANJANA GUPTA}}\\[5pt]
\vspace*{0.5cm}
\includegraphics[height=3cm]{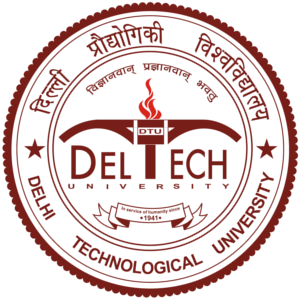}

\vspace*{0.5cm}

{\bf\large DEPARTMENT OF APPLIED MATHEMATICS} \\[5pt]
{\large \mbox{DELHI TECHNOLOGICAL UNIVERSITY}}\\[5pt]
{\large (Formerly Delhi College of Engineering) }\\[5pt]
{\large Bawana Road, Delhi – 110042}\\[10pt]
{\bf \Large May 2023}
\end{center}

\end{titlepage}

\clearpage

\thispagestyle{plain}
\pagenumbering{roman} \setcounter{page}{2}
\begin{center}
\textbf{DEPARTMENT OF APPLIED MATHEMATICS}\\
DELHI TECHNOLOGICAL UNIVERSITY\\
(Formerly Delhi College of Engineering)\\
Bawana Road, Delhi-110042\\
\vspace{1cm}
{\large{\bf{\underline{CANDIDATE’S DECLARATION}}}}
\end{center}

\hfill \break

\begin{spacing}{1.5}
\noindent 
We, (Mansi Yadav) 2K21/MSCMAT/30, (Rashi Jain) 2K21/MSCMAT/40 students of M.Sc. Applied Mathematics, hereby declare that the Project Dissertation titled “The Effect of Covid-19 Lockdown on Human Behaviour Using Analytical Hierarchy Process” which is submitted by us to the Department of Applied Mathematics, Delhi Technological University, Delhi in partial fulfillment of the requirement for the award of the degree of Master of Science, is original and not copied from any source without proper citation. This work has not previously formed the basis for the award of any Degree, Diploma Associateship, Fellowship, or other similar title or recognition.
\end{spacing}

\vspace{4cm}

\noindent Place: Delhi \\
\noindent Date: 27 May 2023 

\begin{flushright}
MANSI YADAV\\
RASHI JAIN
\end{flushright}

\clearpage

\thispagestyle{plain}

\pagenumbering{roman} \setcounter{page}{2}
\begin{center}
\textbf{DEPARTMENT OF APPLIED MATHEMATICS}\\
DELHI TECHNOLOGICAL UNIVERSITY\\
(Formerly Delhi College of Engineering)\\
Bawana Road, Delhi-110042\\
\vspace{1cm}
{\large{\bf{\underline{CERTIFICATE}}}}
\end{center}

\hfill \break

\begin{spacing}{1.5}
\noindent 
I hereby clarify that the Project Dissertation titled “The Effect of Covid-19 Lockdown on Human Behaviour Using Analytical Hierarchy Process” which is submitted by (Mansi Yadav) 2K21/MSCMAT/30, (Rashi Jain) 2K21/MSCMAT/40 Department of Applied Mathematics, Delhi Technological University, Delhi in partial fulfillment of the requirement for the award of the degree of Master of Science, is a record of the project work carried out by the students under my supervision. To the best of my knowledge this work has not been submitted in part or full for any degree or Diploma to this University or elsewhere.
\end{spacing}

\vspace{4cm}

\noindent Place: Delhi \\
\noindent Date: 27 May 2023
 
\begin{flushright}
Prof. Anjana Gupta\\
SUPERVISOR\\
DEPARTMENT OF APPLIED MATHEMATICS\\
DELHI TECHNOLOGICAL UNIVERSITY\\
(Formerly Delhi College of Engineering)\\
Bawana Road, Delhi-110042
\end{flushright}

\clearpage

\thispagestyle{plain}
\begin{center}
{\huge{\bf{ABSTRACT}}}
\end{center}

\hfill \break
\begin{spacing}{1.5}
\noindent 
The coronavirus pandemic corresponds to a serious global health crisis which not only changed the way people used to live but also how people behaved in their daily lives. Information from social and behavioural sciences can help in modifying human behaviour to comply with the recommendations of health officials, as the pandemic requires large-scale behaviour change and puts significant mental stress on individuals. The aim of this paper is to examine the changes in human behaviour brought about by the COVID-19 pandemic, which has caused a global health crisis and altered the way people live and interact. The collection of data has been done through online mode and the behaviour of the people is observed, and the results were finally analysed using the Analytical Hierarchy Process (AHP) which is a multi-criteria decision-making method to rank the factors that had the greatest impact on the changes in human behaviour. During the study, parameters taken under consideration were the ones which were most likely to affect the human behaviour as an impact of COVID-19 lockdown on health, relationship with family and friends, overall lifestyle, online education and work from home, screen time etc. The paper explains each criterion and how it affected human behaviour the most. 

\end{spacing}
\clearpage

\thispagestyle{plain}
\pagenumbering{roman} \setcounter{page}{2}
\begin{center}
\textbf{DEPARTMENT OF APPLIED MATHEMATICS}\\
DELHI TECHNOLOGICAL UNIVERSITY\\
(Formerly Delhi College of Engineering)\\
Bawana Road, Delhi-110042\\
\vspace{1cm}
{\large{\bf{\underline{ACKNOWLEDGEMENT}}}}
\end{center}

\hfill \break

\begin{spacing}{1.5}

It is not possible to outperform without the assistance and encouragement of respective authorities. This one is certainly no exception.
On the very outset of this report, we would like to extend our sincere and heartfelt obligation towards all the personages who have helped us achieve the completion of this dissertation work. Without their active guidance, help, cooperation and encouragement, we would not have made headway in fulfilment of the results.
We are ineffably indebted to Prof. Anjana Gupta for her conscientious guidance and encouragement to accomplish our dissertation.
We both are extremely thankful and pay our gratitude towards each other for coordination and maintaining each one"s individuality on completion of this project. We extend our gratitude to Delhi Technological University for giving us this opportunity.
We also acknowledge with a deep sense of reverence, our gratitude towards our parents
and member of my family, who has always supported us morally as well as economically.
At last but not least gratitude goes to all of our friends who directly or indirectly helped us to complete this project. Any omission in this brief acknowledgement does not mean lack of gratitude.
\\
Thanking You

\end{spacing}

\vspace{2cm}

\begin{flushleft}
MANSI YADAV\\
RASHI JAIN
\end{flushleft}
\clearpage

\tableofcontents

\clearpage
\listoftables
\listoffigures

\newpage

\pagenumbering{arabic}

\setcounter{page}{1}
\mainmatter

\chapter{INTRODUCTION}

The Coronavirus is a first of a sort circumstance at each field. Never before in world's Set of experiences. The 31st December 2019 Covid came into spotlight. On March 24, 2020, the public authority of India requested a cross country lockdown for 21 days, restricting development of the whole populace of 1.3 billion. However, this was presumably an essential, even transient lockdowns, quarantine and social removing can go before long-haul impacts like side effects of mental pressure and confusion, including sleep deprivation, uneasiness, melancholy, and post-horrible pressure side [3]. During this pandemic, protective behaviours include routine hand washing and sanitising, avoiding touching the face, disinfecting phone screens, staying at home when ill, covering up while coughing, eating a balanced diet, keeping physically apart, wearing a face mask, avoiding crowded places, using homoeopathic remedies, putting oneself in quarantine, engaging in digital social networking, and many others. The most frequent emotional reactions are anger, concern, and fear. Also, according to studies Forced closeness is a risk factor for aggression and domestic violence [30]. In this report, we've highlighted a few of the typical human behaviours that emerge during a pandemic crisis. 
Human behaviour
The behaviours, responses, and interactions of people or groups in response to a variety of stimuli, including internal and external aspects including feelings, thoughts, social conventions, culture, and environment, are referred to as human behaviour. A number of interrelated elements, such as genetics, upbringing, education, life experiences, and cultural influences, interact to create human conduct. There are several theories and models that make an effort to explain why people behave the way they do, including:
\begin{enumerate}
    \item Biological theories: These ideas contend that genetic and physiological elements, including hormones, neurotransmitters, and brain structure, have a significant role in determining human conduct.
    \item Behavioural theories: These ideas contend that incentives and punishments, as well as the environment, have a significant influence on how people behave.
    \item Ideas of cognition: These ideas contend that thoughts, beliefs, and attitudes—among other internal processes—are substantially responsible for shaping human conduct.
    \item Social theories: According to these beliefs, social norms, culture, and the influence of others heavily impact how people behave.
    \item Developmental theories: These ideas contend that a person's conduct develops throughout the course of their life and is influenced by a variety of genetic, environmental, and social variables.
\end{enumerate}
Numerous disciplines, such as psychology, sociology, anthropology, economics, and marketing, depend on an understanding of human conduct. It can aid in the development of treatments and tactics to encourage desired behaviour and counteract bad ones, as well as the prediction and explanation of individual and group behaviour.

\section{Impact of Covid-19}
In several nations throughout the world, the COVID-19 epidemic has prompted significant lockdowns and social segregation measures. These policies have significantly changed how people behave, having an effect on relationships with others, economic activity, mental and physical health, and social connections.
Increased social isolation and loneliness are two of COVID-19 lockdowns' most noticeable consequences on conduct. Many people have been cut off from their regular social networks due to limits on meetings and social activities, which has resulted in feelings of loneliness and isolation. Social support is a crucial barrier against mental health issues, therefore this might have a detrimental effect on mental health.
Losses in Business due to Lockdowns have had a severe negative economic impact, causing several businesses to close their doors and countless individuals to lose their employment. For many people and families, this has resulted in financial difficulty, which may have a detrimental effect on mental health and general wellbeing. A person's behaviours and different parts of their life might be significantly affected by losing a business. Depending on the specifics of the company failure, the person's coping style and support system, and their personality, the precise impact might change. Here are a few typical ways that losing a business might alter people's behaviour:
\begin{enumerate}
    \item Psychological effects: Dysfunctional, ashamed, guilty, and depressed sentiments can result from losing a business, which can have a substantial psychological effect on a person. It may also undermine their confidence and sense of self, making it more difficult for them to explore fresh chances.
    \item Financial repercussions: Losing a business may have a big financial impact, since it might mean losing assets, savings, and income. Financial stress and hardship may result from this, which may impact a person's behaviour and cause anxiety, sadness, and unhealthy coping methods including drug misuse.
    \item Losing a business may also have a negative social impact on a person's connections with friends, family, and co-workers. As people may feel embarrassed or humiliated to connect with others, it may cause social isolation and have an adverse effect on their capacity to network and discover new prospects.
    \item Social effects: The stress and anxiety that come with losing a company can also have bodily effects, such as making it harder to fall asleep, changing one's appetite, and raising one's chance of contracting illnesses like high blood pressure, heart disease, and diabetes.
\end{enumerate}

\section{Factors affecting human behaviour}
It has been explored that the Coronavirus pandemic has impacted in general human existence in different ways and the degree of effect is a lot areas of strength for of fundamental areas. Aside from the wellbeing, it has bent the normal life into huge pressure and which results frenzy and dread the whole way across the world.
To assess how the Coronavirus can impact the human conduct an overview (google structure) was planned, specifically the consequences of the examination and determination cycle of a given inquiry, which were filled by various individuals were looked at.
Factors influencing human behaviour:
\begin{itemize}
    \item Physical Characteristics - Age, Health, Illness, Pain, Effect of Substance or Medication
    \item Personal and Intimate Variables - Personality, Beliefs, Attitudes, Emotions, Mental Health
    \item Significant Life Experiences - Family, Culture, Friends, Significant Life Events [8]
\end{itemize}

\begin{figure}[H]
    \centering
    \includegraphics[width=\textwidth, keepaspectratio]{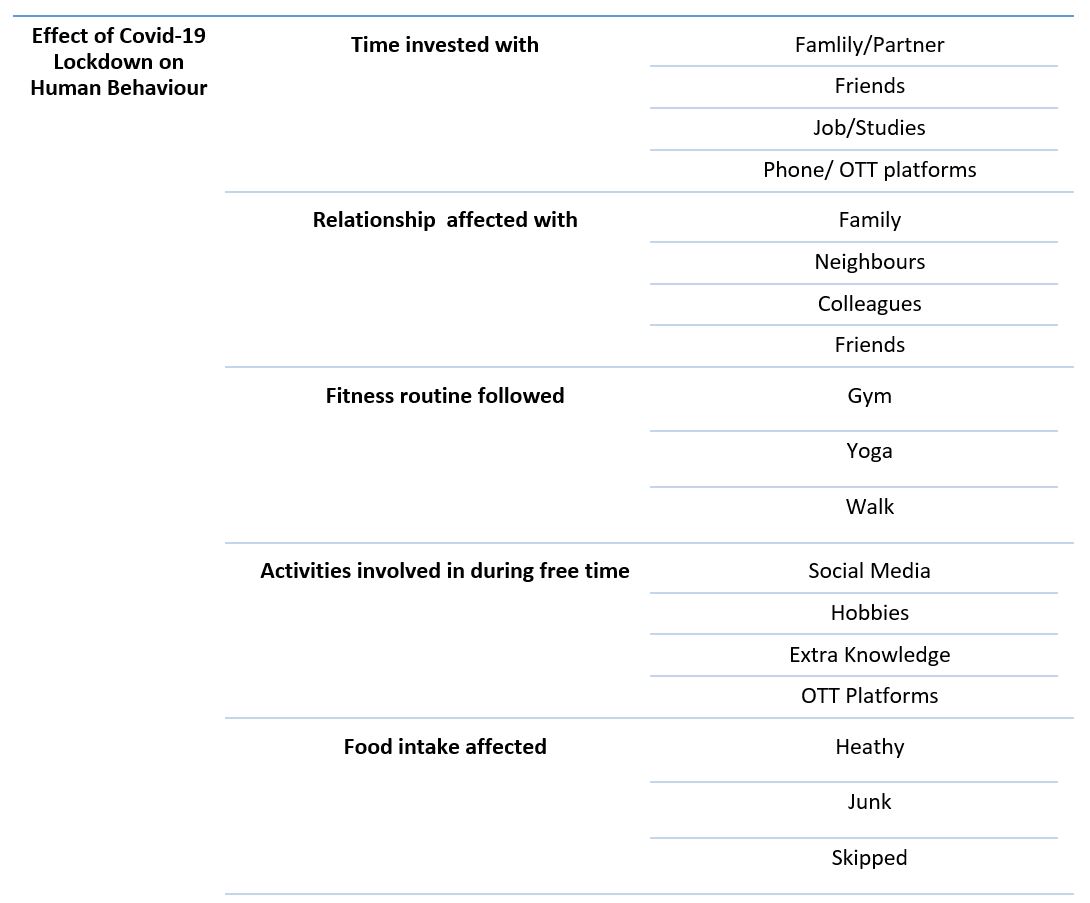}
   \caption{Effect of COVID-19 lockdown on human behaviour}
    \label{fig:my_label}
\end{figure}

While losing a business might be stressful and distressing, it can also be a chance for development and learning, therefore it's vital to remember that. It can assist people in gaining resiliency, flexibility, and fresh talents that they can use in their future endeavours. People may cope with the effects of losing a company by getting assistance from friends, family, and experts like counsellors or mentors. This can help people move on and in the right path.

\section{Effect of lockdown human behaviour}
Lockdown have also affected people's physical health behaviours, making many of them more sedentary and causing access to healthcare to be disrupted. Negative health effects include weight gain, a decline in fitness, and a delay in receiving critical medical care might result from this.
The uncertainty and worry brought on by the epidemic have also increased many people's anxiety and despair. Anxiety and sadness can be exacerbated by the dread of catching the virus, uncertainty about the future, and social isolation. The following are some ways that lockdown and social seclusion have exacerbated anxiety and depression:
\begin{enumerate}
    \item Social isolation: One of the main causes of anxiety and depression is social isolation, which has risen as a result of lockdown and other social isolation techniques. Social engagement is crucial for preserving mental health since humans are social organisms.
    \item Economic stress: Lockdown have also brought about job losses, shortened workdays, and financial uncertainty, all of which can cause stress and worry about money and the future.
    \item Ambiguity: There is a great deal of ambiguity as a result of the epidemic, which can cause emotions of worry and sadness. People are unsure about the duration of the lockdowns, the efficacy of the precautions being taken, and the pandemic's long-term effects on their life.
\end{enumerate}
In general, COVID-19 lockdown have complicated, diverse implications on human conduct, including repercussions on social, economic, mental, and physical health. Decision-makers can prioritise solutions to address the most crucial problems by using the Analytical Hierarchy Process (AHP) to assess the relative relevance of these consequences. AHP can offer insights into the effects of lockdown on human behaviour and suggest areas where focused solutions may be most successful by breaking the problem down into smaller components and giving weights to each component.

\begin{figure}[H]
    \centering
    \includegraphics[scale = 0.5]{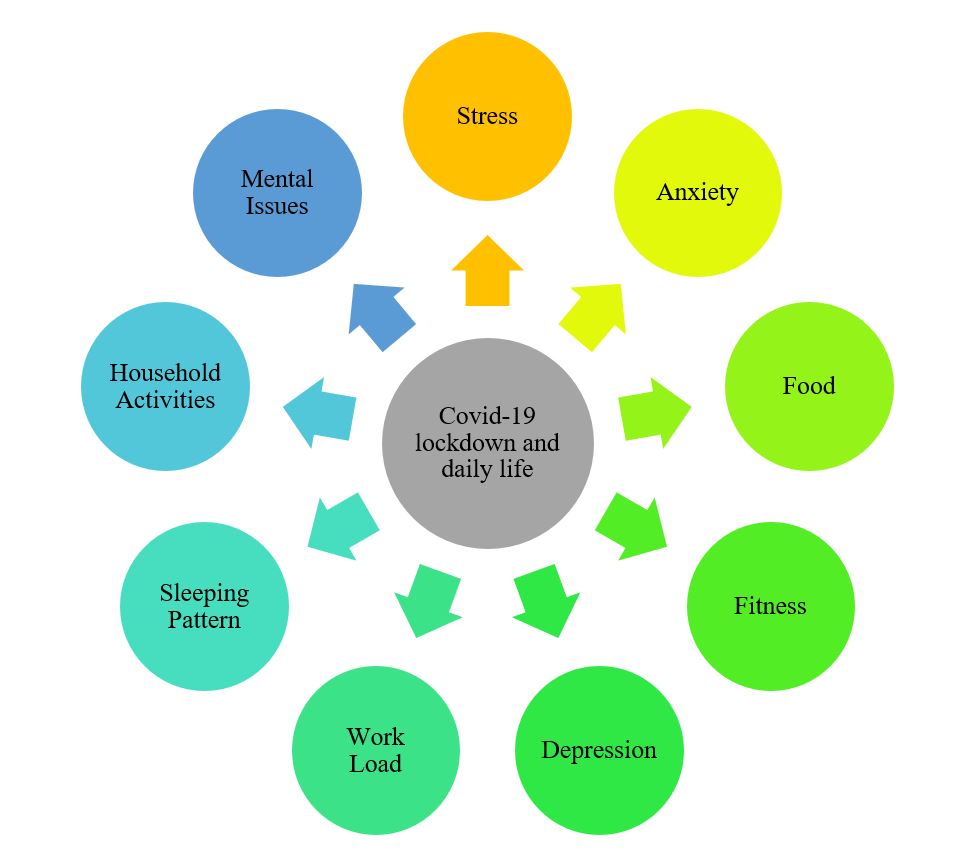}
    \caption{Factors affecting human behaviour during lockdown}
    \label{fig:2}
\end{figure}
The COVID-19 lockdown have had a profound impact on how people behave. To stop the virus from spreading, numerous nations throughout the world have enacted lockdown and social segregation policies. Numerous facets of human conduct, such as social relationships, economic activity, mental and physical health, have been influenced by these metrics.
The rise of social isolation and loneliness is one of the most obvious effects of COVID-19 lockdown on human activity. Many people have been cut off from their regular social networks due to limits on meetings and social activities, which has resulted in feelings of loneliness and isolation. 
To study these various changes in the human behaviour, analytical hierarchy process was taken into account. AHP was created by Saaty [37]. It depends on a progressive construction. It is a useful device for overseeing subjective and quantitative multi-rules components including in dynamic way of behaving. 
A Non-Mathematical and Rational Analysis, the logical pecking order process (AHP) is a cycle that involves progressive decay to manage complex data in multi-measure direction, like data innovation merchant and item assessment what's more, obviously amazing strategies, liberated from any restrictions or presumptions, don't exist for each kind of inquiry one should pose with observational (or even trial) information [9].
AHP philosophy is a basic numerical device that utilizes the idea of pairwise correlation of elements to tackle complex and multi-layered issues by producing a various leveled structure. The idea of AHP was first used to find an answer connected with military for arranging and asset division and is as of now utilized in various spaces like wellbeing, designing, training, industry, and others [14]. In this strategy the overall significance or loads of tried factors are resolved which helps in positioning the variables which helps in pursuing exact choices. The advantages of AHP technique over other MCDM strategies are its gracefulness, instinctive enticement for policymakers and its capacity to guarantee predictable outcomes [10]. It uncovered various profound results, including pressure, wretchedness, crabbiness, a sleeping disorder, dread, disarray, outrage, dissatisfaction, weariness, and disgrace related with isolation, some of which continued after the quarantine was lifted [27].
The purpose of this paper is to determine which variable has the greatest impact on human behaviour. To do this, the Analytic Hierarchy Process (AHP) is used to choose the appropriate factor. AHP has been successfully used in various applications, including selecting the best option from multiple alternatives. Factor selection is relevant to human behaviour, so the use of AHP is justified. The advantage of this selection tool is that it is based on a pairwise comparison of the options and the user's preferences.  The authors aim to evaluate the effectiveness of alternative methods for evaluating pairwise comparisons. By adjusting one element in the pairwise comparison network (while maintaining others constant), the factor selection process can be improved and modifications can be made [33].

\chapter{LITERATURE REVIEW}
Using protective measures when there is neither a vaccination nor a proven cure for the new coronavirus. Between nations and continents, there are differences in incidence and case fatality rates. Due to this, the chance of developing the disease is quite unknown.  Individual coping methods allow people to deal with health issues through various information practises, while a variety of information practises, experiences, and communication channels help people increase their capacity to cope in the setting of sickness [23]. Several studies that study the sudden shift to online teaching and learning activities during the early months of the pandemic highlight to information issues, such as supporting students' digital skills, particularly from the perspective of students, notably in university environments [11]. Many of the studies dealt with theoretical viewpoints or analyses of fake news documents rather than actual people's behaviour, giving the impression that misinformation is a less developed field of study. Additionally, user interactions with digital health tools have been studied to better understand how they might enable remote care [15].
WHO articulated the eruption to be an overall prosperity emergency of worldwide concern. The Emergency Board has given advice to WHO, to People's Republic of China, to all countries, and to the overall neighbourhood, measures to control this episode. The Warning gathering acknowledges that interrupting infection is at this point possible spread, considering that countries set solid areas for up to recognize affliction early, independent and treat cases, follow contacts, and advance social eliminating measures proportionate with the bet [35]. We distinguish a few bits of knowledge for powerful reaction to the Coronavirus pandemic and feature significant holes specialists ought to move rapidly to fill before long also, months [27]. 

\begin{table}[H]
\centering
\resizebox{\columnwidth}{!}{%
\renewcommand{\arraystretch}{1.5}
\begin{tabular}{lllllllll}
\hline
\textbf{Author}             & \textbf{Covid-19} & \textbf{Human behaviour} & \textbf{Mental Health} & \textbf{Physical Fitness} & \textbf{Work} & \textbf{Relationships} & \textbf{Social Media} & \textbf{AHP} \\ \hline
\text{[9]}       &                   & \checkmark                        &                        &                           &               &                        &                       & \checkmark             \\
\text{[8]} &                   & \checkmark                       & \checkmark                       & \checkmark                         & \checkmark              &                        & \checkmark                      &              \\
\text{[2]}        & \checkmark                  & \checkmark                        &                        &                           &               &                        & \checkmark                     &              \\
\text{[27]} & \checkmark                & \checkmark                       & \checkmark                     & \checkmark                          &               &                        &                       &              \\
\text{[18]}       & \checkmark                  & \checkmark                        & \checkmark                       & \checkmark                         & \checkmark              & \checkmark                       & \checkmark                      &              \\
\text{[23]}               & \checkmark                 & \checkmark                      & \checkmark                      &                           &               &                        & \checkmark                     &              \\ \hline
\textbf{This research}      & \checkmark                & \checkmark                        & \checkmark                      & \checkmark                   & \checkmark             & \checkmark                       & \checkmark                      & \checkmark             \\ \hline
\end{tabular}%
}
\caption{Table of Authors}
\label{Table of Authors}
\end{table}

Uncertain prognoses and impending severe resource constraints for testing, treatment, and safeguarding first responders and the healthcare system. Among the main stressors that undoubtedly will contribute to widespread emotional distress and an increased risk for psychiatric illness associated with Covid-19 are providers from infection, imposition of unfamiliar public health measures that infringe on personal freedoms, large and growing financial losses, and conflicting messages from authorities. As part of the pandemic response, health care professionals play a crucial role in managing these emotional consequences. Public health emergencies may have an impact on people's health, safety, and wellbeing on an individual level, for instance, insecurity, confusion, emotional isolation, and stigma) as well as on a community level (due to economic loss, closures of businesses and educational institutions, insufficient funding for medical assistance, and insufficient distribution of basic necessities)[3].
It has been explored that the Coronavirus pandemic has impacted generally human existence in different ways and the degree of effect is a lot areas of strength for of principal areas. Aside from the wellbeing, it has curved the standard life into enormous pressure and which results frenzy and dread all over the world. The COVID-19 outbreak has been declared a public health emergency of global concern by the WHO. The WHO is dedicated to carrying out a comprehensive risk communication strategy to stop the spread of COVID-19, improve public health protocols for containing the present outbreak, protect the healthcare workers and the health system's resilience, and expand investigations into cases across China. To forestall the spread of Coronavirus the idea like remaining at home and social separating were generally taken on around the world.  Within excess of 80 nations shutting their lines and requesting them to close down their specialty unit. Individuals across the world began dropping their vacation and work excursions, this straightforwardly impacted the inn business, occasion industry, travel industry, aircraft industry and different ventures which were straightforwardly or by implication related with them [1],[18]. Specialists and government offices are cautioning about another emergency for example emotional well-being issues, despondency, uneasiness and ascend in self-destruction rates that will outlast the Covid pandemic and can endure up to weeks, months and even years. Patients who have COVID-19 infection are more likely to experience a wide range of psychological effects, and this illness may have a significant impact on aged, parenting, relationships, marriage, and family links [32]. 
Coronavirus pandemic has caused feeling of confinement due to social separating, feeling of dread toward disease and demise about themselves as well as friends and family, vulnerability about future, disturbances in everyday daily schedule, changes in dietary and rest propensities, denied get-togethers. Lockdown and social removing standards are compelling individuals to remain inside, this can make a pessimistic impact on the actual wellbeing and way of life of the majority particularly individuals living in metropolitan regions. Absence of actual work, stationary way of life, less time spent in sports exercises and exercise because of conclusion of centres and arenas, sporadic and expanded resting time, gorging and admission of unequal food because of inaccessibility of specific food sources can prompt expanded adiposity levels in this way bringing about corpulence and other medical problems. The results highlight detrimental changes in daily routine after a brief time. Findings allow for the establishment of suggestions for maintaining healthy behaviours with the fewest detrimental health effects under similar pandemic settings. This is especially true of nations like Italy and Spain a serious danger and more stringent measures [16].
Since the pandemic has made face-to-face interaction rare, children's technology use has increased much beyond those figures. In actuality, it has almost doubled. Prior to this catastrophe, youngsters of all ages spending, on average, only around 3 hours per day using screens. Concerns have been raised concerning the potential negative impacts of excessive screen time on youngsters as a result of this abrupt and major increase. 
With the majority of the youngsters telecommuting, internet advancing as an option in contrast to study hall learning, conclusion of films, shopping centres and other amusement sources individuals will generally utilize screens more frequently than they regularly do either to go to proficient obligations, to engage themselves by buying into web stations, covering service bills on the web or for remaining associated with companions and family members through virtual entertainment. Unreasonable screen time additionally presents wellbeing perils like mind harm, deficiency of consideration and discernment control, eye fatigue, opportunities to foster near sightedness and other vision issues [36].

\chapter{METHODOLOGY}

The study aimed to assess the effects of the COVID-19 pandemic on various aspects of society, including physical and mental health, the economy, misinformation spread, online education, relationships with loved ones, usage of social media, views on lockdown and social distancing measures, and other factors. These aspects include physical health, mental health, economic health, Infodemics, online education, relationships with family and spouse, usage of social media, perception of the lockdown, social distancing as a measure to restrict the spread of COVID-19 infection, etc. The goal was to gain insights into people's experiences and perspectives during the pandemic and explore potential solutions to the global crisis caused by COVID-19.
The analytic hierarchy process (AHP) is an effective method for quantitative analysis that was introduced in the early 1977s, by American scientist, Professor Saaty of Pittsburgh University [29]. 
The Analytic Hierarchy Process (AHP) matrix is a decision-making tool that makes it easier to compare two criteria or options side by side. Each member in the matrix indicates the relative relevance of one criteria or option compared to another, and the matrix is used to assign values to the pairwise comparisons. The square AHP matrix has rows and columns that stand in for the different comparison criteria or options. The diagonal matrix elements, which indicate the comparison of an element with itself, are always given a value of 1. The pairwise comparisons are represented by the off-diagonal components, which are given values based on the relative relevance of the elements being compared. 
The Analytic Hierarchy Process uses the AHP matrix as a tool to rank pairwise comparisons of criteria or alternatives. The weights of the criteria or alternatives are determined by the matrix and then evaluated and prioritised according to the decision-maker's priorities. The AHP matrix offers an organised and methodical approach to decision-making, which can aid in ensuring that choices are supported by reasonable justification and impartial standards.
The Analytic Hierarchy Process (AHP) is a systematic decision-making approach that entails segmenting complicated issues into smaller parts and giving each part a proportional weight. The steps in the methodology are as follows:

\begin{enumerate}
    \item Identify the issue: The AHP's first phase is to identify the issue and the decision's goals. It is crucial to describe the issue and the selection criteria precisely so that they may be used to the evaluation of potential solutions.
    \item Create a structure: Next, a hierarchy is created to divide the issue into more manageable parts. The hierarchy is made up of a goal, standards, and options. The aim is the decision's overarching purpose, the conditions are the parameters by which the alternatives will be assessed, and the options are potential fixes for the issue.
    \item Pairwise comparisons: Following the development of the hierarchy, pairwise comparisons are performed to ascertain the relative significance of each component. Pairwise comparisons compare every element in the hierarchy against every other element to decide which is more significant. On a scale from 1 to 9, where 1 denotes equal importance for both components and 9, excessive importance for one component, the comparisons are done.
A square matrix \(A=a_{ij} \) where \(1\leq i,j \leq n\) and 
\[a_{ij}\in \{9,8,7,6,5,4,3,2,1,1/2,1/3,1/4,1/5,1/6,1/7,1/8,1/9\}\]
\(n\) is the number of pairs compared. The number of judgements required in formation of the symmetric matrix is  \(n(n-1)/2\). The elements on the diagonal of the matrix are equal to 1. 
The primary principle of AHP is to compare two elements—\(i\) on the left side of the matrix and j on top—and determine which one has the quality or meets the criterion more, i.e., which one is considered more significant under that criterion and by how much. The target respondents were requested to rate the factors related to COVID-19. From the survey results conducted, we took the geometric mean of all the alternatives, and then substituted them in the upper triangular pairwise comparison AHP matrix. 
For ranking, a 9-pointer scaling system is used. The table shows the relationship between \(i\) and \(j\).

\begin{table}[H]
\centering
\renewcommand{\arraystretch}{1.2}
\resizebox{\columnwidth}{!}{%
\begin{tabular}{c|l|l}
\hline
\textbf{Intensity of Importance} & \textbf{Definition}                                                                                                                                                                                      & \textbf{Explanation}                                                                                                                        \\ \hline
1                                & Equal importance                                                                                                                                                                                        & \begin{tabular}[c]{@{}l@{}}Two activities contribute equally to the \\ objective.\end{tabular}                                              \\ \hline
3                                & Weak importance of one over another                                                                                                                                                                     & \begin{tabular}[c]{@{}l@{}}Experience and judgment slightly lean towards \\ one activity being more favourable than the other.\end{tabular} \\ \hline
5                                & Essential or strong importance                                                                                                                                                                          & \begin{tabular}[c]{@{}l@{}}Experience and judgment significantly favour\\ one activity over the other.\end{tabular}                         \\ \hline
7                                & Very strong importance                                                                                                                                                                                  & \begin{tabular}[c]{@{}l@{}}An activity is strongly favoured and its \\ dominance demonstrated in practice.\end{tabular}                     \\ \hline
9                                & Absolute importance                                                                                                                                                                                     & \begin{tabular}[c]{@{}l@{}}The evidence favoured one activity over another\\  is of the highest possible order of affirmation.\end{tabular} \\ \hline
2,4,6,8                          & \begin{tabular}[c]{@{}l@{}}Intermediate values between the two \\ adjacent judgments\end{tabular}                                                                                                       & When to make a compromise.                                                                                                                  \\ \hline
Reciprocals of above, nonzero    & \begin{tabular}[c]{@{}l@{}}If activity i has one of the above nonzero \\ numbers assigned to it when compared with \\ activity j, then j has the reciprocal value \\ when compared with i.\end{tabular} &                                                                                                                                             \\ 
\end{tabular}%
}
\caption{9 pointer scale used for scaling}
\label{4}
\end{table}

\item Calculate weights: Following pairwise comparisons, each component's weights are determined. The pair-wise comparison scores are taken into consideration when calculating the weights using a mathematical procedure. The weights show how significant each element is in relation to the others in the hierarchy.
\item Verify for consistency: To make sure that the weights are precise and dependable, it is crucial to verify for consistency in the pair-wise comparisons. For each component, a consistency ratio (CR) is computed to accomplish this. By comparing the actual weights to the anticipated weights based on the pairwise comparisons, the CR is determined. Pairwise comparisons are regarded as consistent if the consistency ratio is below 0.1.
The model supports a pairwise comparison approach to describe the relative importance of the performance Criteria and assign weights to them according to their importance. Pairwise comparisons are computed using a scale. Such a scale is an injective mapping between the set of discrete linguistic choices available to the decision maker and a discrete set of numbers which represent the importance, or weight, of the previous linguistic choices. The scale proposed by Saaty is described \ref{Table 3}

\begin{table}[H]
\centering
\setlength\tabcolsep{13pt}
\begin{tabular}{c|lllllllll}
\hline
 \(n\) & 1 & 2 & 3    & 4    & 5    & 6    & 7    & 8    & 9    \\ \hline
 \(RC\) & 0 & 0 & 0.58 & 0.89 & 1.12 & 1.24 & 1.32 & 1.14 & 1.45 \\
\end{tabular}
\caption{Random Consistency Index (RC)}
\label{Table 3}
\end{table}

\item Evaluate alternatives:  Alternatives can be assessed based on the criteria and their relative weights after the weights have been determined. For each criterion, each alternative is given a score, and the scores are multiplied by the weights to provide the alternative's total score.
\item Sensitivity analysis: The decision's robustness is examined in the final step of the process. To do this, adjust the weights of the criteria and see how the final choice is affected. Sensitivity analysis can assist in determining the most crucial factors and the effect of ambiguity on the choice.
\end{enumerate}
In conclusion, utilising AHP to examine how COVID-19 lockdowns affect people's conduct might assist decision-makers choose solutions to address the most pressing problems while also offering a formal and methodical approach to understanding the complicated subject. AHP provides for a more nuanced understanding of the effects of lockdowns on human behaviour by breaking the problem down into smaller components and allocating weights to each component, and it can reveal areas where focused solutions may be most successful.
The study may, for instance, show that the influence on mental health is the most crucial element, with sub-components like anxiety and depression being especially relevant. This might help to lessen the detrimental impact of lockdowns on mental health by informing the creation of mental health therapies like teletherapy or online support groups. 
Similar to this, the study may show that the influence on economic activity is a key issue, with special importance given to its sub-components like job loss and financial hardship. This may help in the creation of economic interventions to support those impacted by lockdowns, such as small company loans or financial aid programmes.
In conclusion, employing AHP to examine how COVID-19 lockdowns affect people's conduct is a useful tool for decision- and policy-makers. AHP can assist identify the most significant aspects and influence the creation of focused actions to lessen the negative effects of the pandemic by offering an organised and methodical approach to comprehending the complicated situation.

\chapter{DISCUSSION AND RESULTS}

\section{Data Overview}
During COVID-19 lockdown, information was gathered using a questionnaire created to assess many facets of human behaviour. People from the general public who represented a range of ages, occupations, and geographical areas were among the responses. In order to understand how COVID-19 has affected people's daily lives, mental and physical health, relationships with family, screen time, etc., the survey questionnaire included both self-reports and closed-ended multiple choice questions about COVID-19. Based on the qualitative information collected from the participants, this study presents the major findings.
In this study, the AHP method was used; 238 participants were selected. All participants were highly qualified in their domain of study. A questionnaire was answered by them, and they presented their views on change in their behaviour during the lockdown because of COVID-19, they were asked to score and rank the factors. The respondents were invited to complete the questionnaire based on their behavioural, phycological, and physiological changes. Out of 238 participants, 94 (39.49 \%) were males and 144 (60.50\%) were females. Majority of 200 youngsters in the ages 18-35 years (84.03\%), 35-60 years (13.44\%) and remaining children and senior citizens (2.94 \%). All responses were anonymous and no economic compensation was presented. 
To make the results more precise and quantitative, the factor variables were arranged. Data were gathered using the survey tool (questionnaire). Data was gathered from various geographic regions, and the survey instrument was delivered online. Data collection took place between October 2022 and January 2023. All of the gathered data was correctly categorised, cleansed, and then analysed without any alterations. The analysis' material provides the probability anticipated behaviour for COVID-19. 

\begin{table}[H]
\centering
\setlength\tabcolsep{29pt}
\begin{tabular}{lcc}
\hline
                           & \textbf{Total Respondents} & \textbf{\%age}     \\ \hline
 \textbf{Age}               &                   &           \\ \hline
1-18                       & 5                 & 2.1       \\
18-35                      & 200               & 84.03     \\
35-60                      & 31                & 13.44     \\
60+                        & 2                 & 0.84      \\ \hline
\textbf{Employment Status} & \textbf{}         & \textbf{} \\ \hline
Student                    & 156               & 65.54     \\
Employment                 & 61                & 25.63     \\
Unemployed                 & 12                & 5.04      \\
Others                     & 9                 & 3.78     \\ \hline
\end{tabular}
\caption{Characteristics of the sample}
\label{Table 2}
\end{table}

\noindent The COVID-19 pandemic has caused a significant loss of human life and poses a significant challenge to overall health, food systems, and the world of work. The financial and social disruption caused by the pandemic is devastating. In India, the current number of active cases is 1768, as reported by the Ministry of Health and Family Welfare [22].
In our survey it was found that 19\% tested positive, 26\% tested negative, 19\% reported symptoms but were not tested, and only 36\% believed they were not infected and had no symptoms.

\begin{figure}[H]
    \centering
    \includegraphics[scale = 0.5]{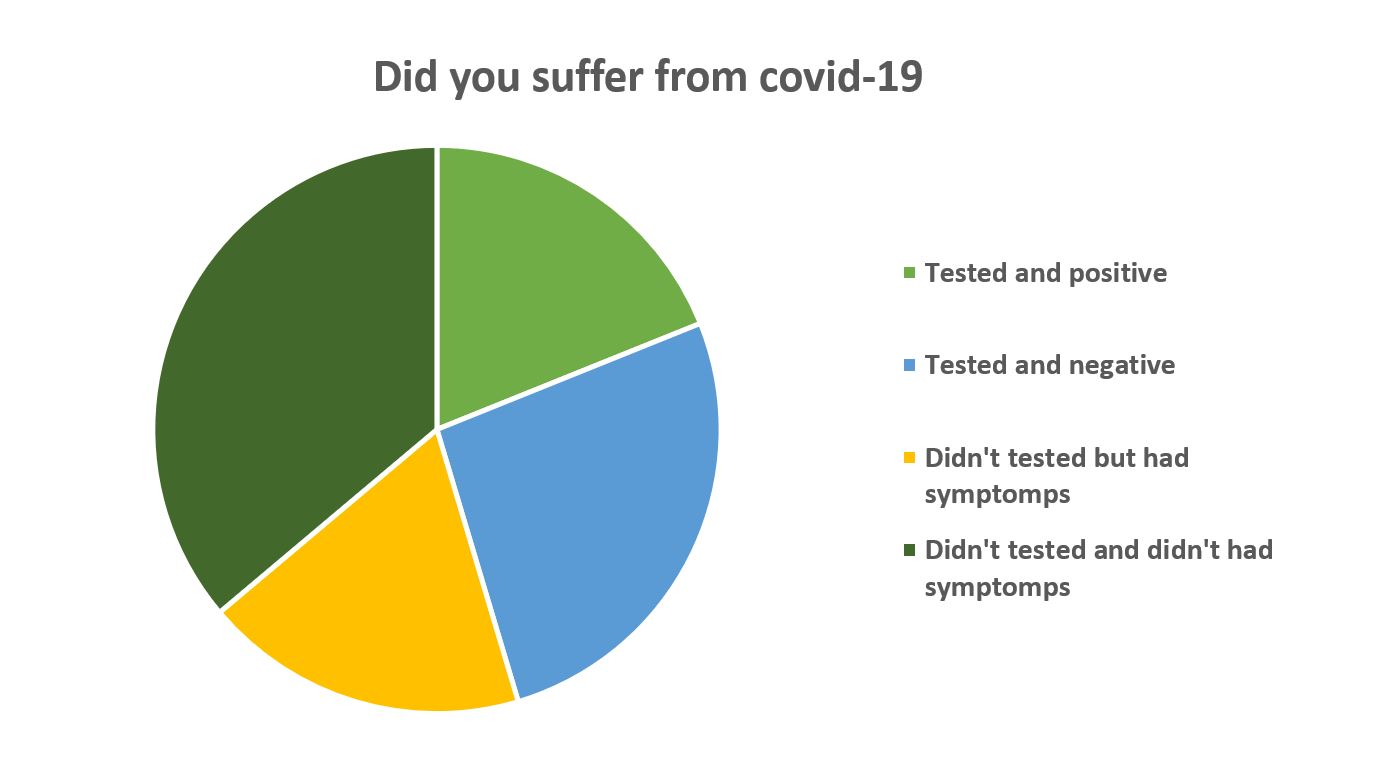}
    \caption{Did you suffer from COVID-19}
    \label{fig:3}
\end{figure}

A study found that the coronavirus pandemic caused significant mental distress, including anxiety, stress, and depression, among Chinese nationals [31]. Another study conducted among Chinese nationals also found that stress, anxiety, and depression were prevalent and affecting individuals during the pandemic [28]. 
In our survey, 30\% of the respondents replied positively, while 70\% replied negatively when asked about their mental problems. 70\% reported experiencing stress sometimes, 21\% reported not experiencing any stress, and 9\% reported experiencing stress most of the time. When asked about their feeling of downheartedness 60\% of the respondents reported feeling down at some point during the lockdown, while 16\% reported never feeling down and 24\% reported not facing such issues. 

\begin{figure}[H]
    \centering
    \includegraphics[width=\textwidth, keepaspectratio]{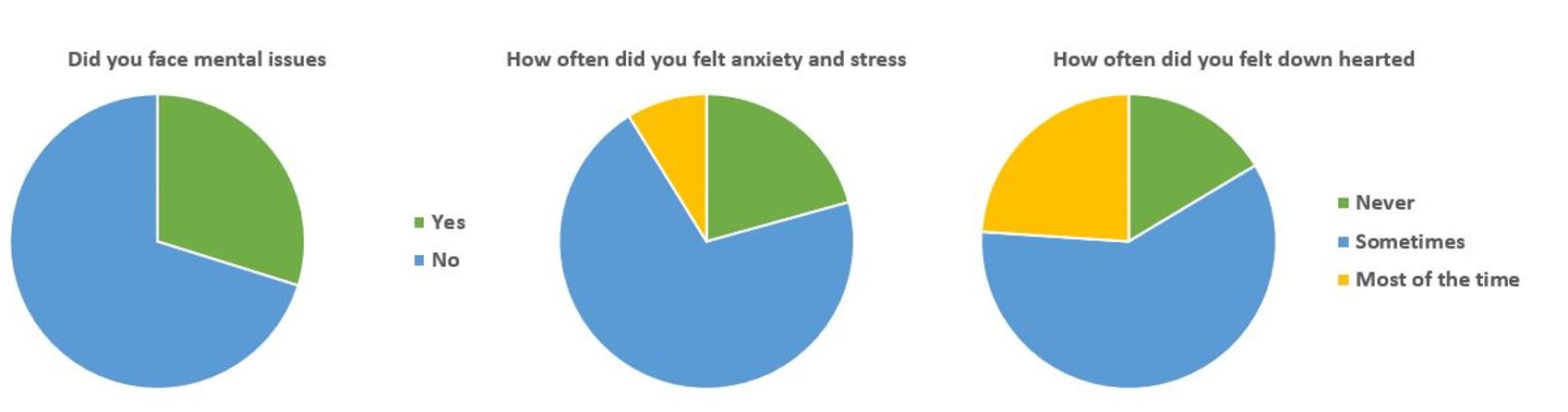}
    \caption{Related to Mental issues}
    \label{fig:4}
\end{figure}

The outright contrasts between the availability of food during the quarantine and pre-quarantine we can take note of that the recurrence of labels like Bread, Flapjack/Tortilla/Oatcake, Cheddar, Flour, Spread, Baked good, Eggs, Sauce/dressing, has expanded during the quarantine, while for semantic labels, for example, Fish, Citrus natural product, Salt,  Fat/oil, Corn/cereals/grains, Sugar (syrup/honey/chocolate), Organic product juice/squash, has diminished [13].
When people were asked about their food habits during the lockdown, majority (60\%) reported that they never skipped a meal, 20\% reported skipping one meal per day, and 20\% reported reducing food intake once a week or 1-2 times a month. Regarding unhealthy food, 29\% reported never eating junk food, 33\% reported eating it once a week, and 38\% reported eating it 1-2 times a week or daily. When it comes to healthy food, majority (86\%) reported eating it every day, while 14\% reported otherwise.
Millions of people around the world experienced significant daily and routine changes as a result of the lockdown imposed in the wake of the COVID-19 pandemic in spring 2020. With the use of a cutting-edge survey of lockdown families from Italy, the United Kingdom, and the United States, we examine how these changes affected the patterns of activity within the home. A significant portion describe changes in family routines, including different job schedules, chore assignments, and tensions within the home. Despite the fact that more men are helping with childcare and food shopping, re-allocations are not as drastic as interruptions to work schedules might indicate, and families that have to redistribute responsibilities report higher tensions [4].

\begin{figure}[H]
    \centering
    \includegraphics[width=\textwidth, keepaspectratio]{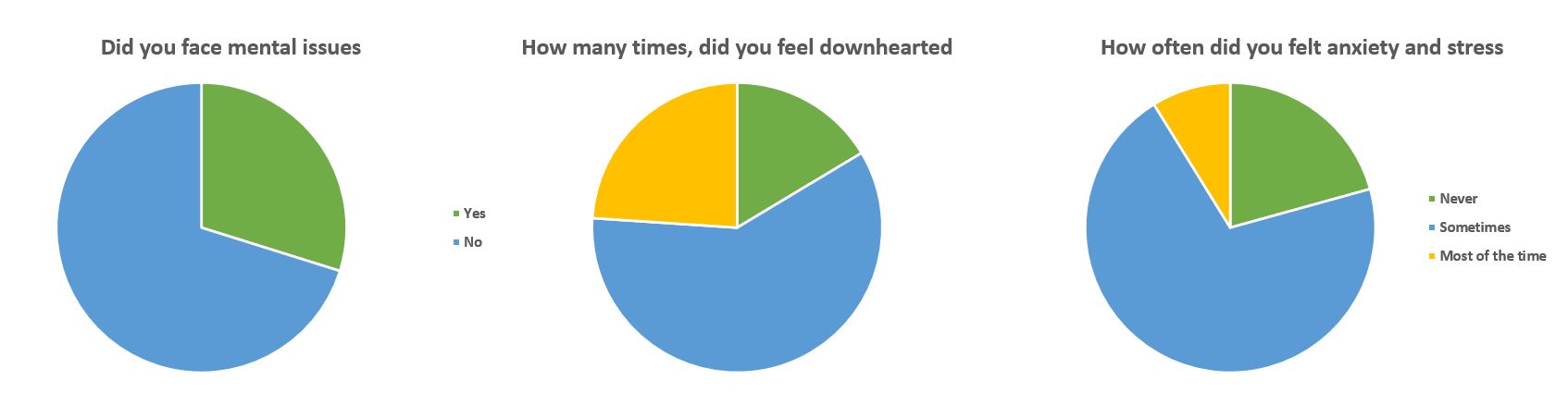}
    \caption{Food habits during COVID-19 lockdown}
    \label{fig:5}
\end{figure}
When asked about time spent on family tasks such as laundry, dish washing, and home cleaning, 67\% of the respondents reported spending 5-7 hours on these activities, 23\% reported spending 8-10 hours, and 10\% reported spending 0-4 hours.
\begin{figure}[H]
    \centering
    \includegraphics[scale = 0.5]{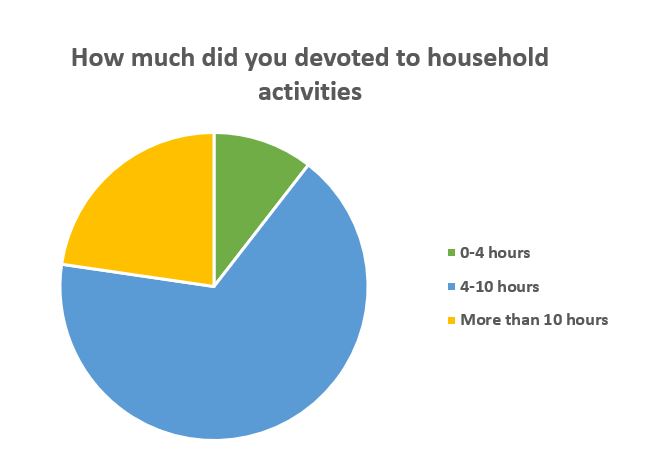}
    \caption{Time devoted to household activities}
    \label{fig:6}
\end{figure}
By limiting physical activity and exposure to daylight, the lockdown restriction might have an adverse effect on health and wellness. Social isolation may also raise stress levels. These modifications may have an effect on a person's daily activities, circadian rhythm, and sleep cycle [5].
\begin{figure}[H]
    \centering
    \includegraphics[scale = 0.5]{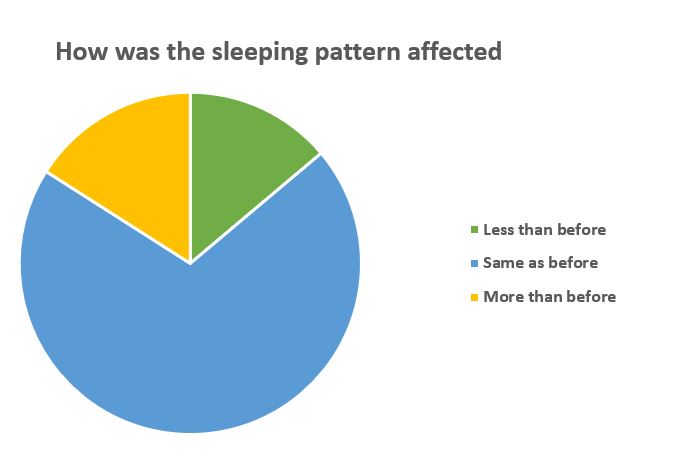}
    \caption{Sleeping pattern affected}
    \label{fig:7}
\end{figure}
In general, but not always, stress and sleep quality are antithetical. Sleep reactivity affects how stress affects sleep quantity, quality, and duration [19]. Also, the use of digital media before bedtime rose throughout the lockdown, and people's sleep patterns changed noticeably, with later bedtimes and reported poorer sleep quality.
When the respondents were asked about changes in their sleep patterns during the COVID-19 lockdown, 14\% reported sleeping for 0-4 hours, 70\% reported sleeping for 5-7 hours, and 16\% reported sleeping for 8-10 hours.

\section{Analytical Hierarchy Process}

\subsection{Time spent on average}
Our first analysis focused on the amount of time people spent with their friends and family, as well as in their surroundings, before and during the COVID-19 pandemic. Before the pandemic, people generally divided their time between friends, family, and work. However, during the lockdown, it was interesting to observe how people allocated their time to their loved ones.
Spending time with friends and family is a common and important aspect of many people's lives. It helps to maintain relationships, foster connections, and promote overall well-being [38]. People may participate in various activities with their friends and family, such as dining out, playing sports, or simply spending time together at home. However, the onset of the COVID-19 lockdown severely impacted these gatherings and meetups. As a result, some people had the opportunity to spend more time with their families, while others were overwhelmed with work and unable to spend as much time with their loved ones.
Additionally, people started to spend more time on their mobile phones, and it was important to understand how they were spending most of their day. With the shift to online work and education [7], multinational companies and schools were affected, leading to an increase in the amount of time people spent on their jobs or studies.

\begin{table}[H]
\centering
\setlength\tabcolsep{10pt}
\renewcommand{\arraystretch}{1.2}
\begin{tabular}{l|cccc}
\hline
\textbf{Criteria}       & \textbf{Friends} & \textbf{Family/Partner} & \textbf{Phone} & \textbf{Job/Studies} \\ \hline
\textbf{Friends}        & 1.0000           & 0.2857                  & 0.3750         & 0.4286               \\
\textbf{Family/Partner} & 3.5000           & 1.0000                  & 1.0000         & 1.0000               \\
\textbf{Phone}          & 2.6667           & 1.0000                  & 1.0000         & 0.7149               \\
\textbf{Job/Studies}    & 2.3334           & 1.0000                  & 1.4000         & 1.0000              
\end{tabular}%
\caption{Pair-wise comparison matrix for time spent on an average}
\label{5}
\end{table}

In the relative importance assessment of the two elements, reciprocal axiom is applied, meaning that if the element \(i\) is assessed 6 times more important than \(j\) then must be 1/6 times the significance rather than element \(i\)[39]. Moreover, the comparison of the same two elements will result in a number 1, meaning equally important, here friends are equally important to friends.
The largest left eigenvector for a judgement matrix with pairwise comparisons is approximated by taking the geometric mean of each row. That is, the elements in each row are multiplied with each other and then the \(n^{th}\) root is taken (where n is the number of alternatives considered). Here the geometric mean we get by calculating \((1 \cdot \frac{2}{7} \cdot \frac{3}{8} \cdot \frac{3}{7})^\frac{1}{4}\)= 0.4629. So, the priority of "Friends" is 0.4629, according to the geometric mean. In the same way, we get the geometric mean of the rest of the rows "Family" as 1.3678, “Phone” as 1.1748 and "Job" as 1.3444.
\begin{table}[H]
\centering
\renewcommand{\arraystretch}{1.2}
\setlength\tabcolsep{6pt}
\begin{tabular}{l|ccccc}
\hline
\textbf{Criteria}       & \textbf{Friends} & \textbf{Family/Partner} & \textbf{Phone} & \textbf{Job/Studies} & \textbf{PV} \\ \hline
\textbf{Friends}        & 1.0000           & 0.2857                  & 0.3750         & 0.4286               & 0.4629      \\
\textbf{Family/Partner} & 3.5000           & 1.0000                  & 1.0000         & 1.0000               & 1.3678      \\
\textbf{Phone}          & 2.6667           & 1.0000                  & 1.0000         & 0.7149               & 1.1748      \\
\textbf{Job/Studies}    & 2.3334           & 1.0000                  & 1.4000         & 1.0000               & 1.3444      \\ \hline
                        &                  &                         &                & \textbf{Sum}         & 4.3499     
\end{tabular}%
\caption{Calculation of the priority vector}
\label{6}
\end{table}

Once the pairwise comparisons are done, they are then synthesized to determine the priority. For this purpose, the Normalization of the Geometric Mean (NGM) is taken into account (Hsiao, 2002). It is done by dividing each priority value by the sum of all priorities that we got from the geometric mean. For doing so, all priority values are added, 

\begin{equation}\label{eq:1}
    (0.4659+1.3678+1.1748+1.3444=4.3499)
\end{equation}

Each priority value is then divided by the number 4.3499 as obtained in \ref{eq:1} . In that way, we get the final normalized values for "Friends" as 0.1064, "Family" as 0.3144, “Phone” as 0.2701 and "Job" as 0.3091.
We obtain the Principal Eigen Value from the pairwise comparison matrix after calculating the priority vector. We now execute a matrix multiplication, multiplying the priority vector by the pairwise comparison matrix, in order to obtain the principal eigenvalue. The result of matrix multiplication is divided by the cell value of the related priority vector. Then, the consistency measure is obtained by the average of this resulting vector, which is 
\begin{equation}\label{eq:2}
    \frac{4.0406+4.0263+4.0324+4.0442}{4}=4.0359
\end{equation}

\begin{table}[]
\resizebox{\columnwidth}{!}{%
\renewcommand{\arraystretch}{1.3}
\begin{tabular}{l|ccccccccc}
\hline
\textbf{Criteria}       & \textbf{Friends} & \textbf{Family/Partner} & \textbf{Phone} & \textbf{Job/Studies} & \textbf{PV} & \textbf{NW} & \textbf{EV}      & \textbf{CM} &  \\ \hline
\textbf{Friends}        & 1.0000           & 0.2857                  & 0.3750         & 0.4286               & 0.4629      & 0.1064      & 0.4299           & 4.0406      &  \\
\textbf{Family/Partner} & 3.5000           & 1.0000                  & 1.0000         & 1.0000               & 1.3678      & 0.3144      & 1.2660           & 4.0263      &  \\
\textbf{Phone}          & 2.6667           & 1.0000                  & 1.0000         & 0.7149               & 1.1748      & 0.2701      & 1.0891           & 4.0324      &  \\
\textbf{Job/   Studies} & 2.3334           & 1.0000                  & 1.4000         & 1.0000               & 1.3444      & 0.3091      & 1.2499           & 4.0442      &  \\ \hline
                        &                  &                         &                & \textbf{Sum}         & 4.3499      & 1           & \textbf{Average} & 4.0359      & 
\end{tabular}%
}
\caption{Calculation of Eigenvectors and consistency}
\label{7}
\end{table}
Where, PV=Priority Vector, NW=Normalized Vector, EV=Eigen Vector and CM=Consistency Measure.
One of the most real-world problems in the AHP methodology is that it permits slightly non-consistent pairwise comparisons. As sample size is a factor, it's important to note that AHP is a subjective method for addressing specific problems. Therefore, surveys conducted using this method do not necessarily require a large sample size. In fact, a large sample size may lead to a higher degree of inconsistency. The literature on AHP surveys suggests that small sample sizes are more appropriate and reliable when focusing on the decision problem being studied. This is because the likelihood of receiving arbitrary feedback decreases with a small sample size, resulting in a higher degree of consistency. In case all the comparisons are consistent perfectly, then \(a_{ij}=a_{ik}\cdot a_{kj}\) is always true for any combination of comparisons taken from the judgment matrix. However, perfect consistency is rare in practice. This capability of the AHP to absorb a certain level of inconsistency is extremely valued. If the corresponding consistency ration (CR), which is the measurement of how much the results have deviated from the consistency, is less than 10\%, then in the AHP method, the pairwise comparisons are considered to be adequately consistent, otherwise, it needs to be reviewed. For this, first of all, estimation of the consistency index (CI) is done as given in \ref{eq: 3}, by adding the columns in the judgment matrix and multiplying the resulting vector by the priority vector, calculated before. As a result of this, maximum eigen value is obtained, denoted by \(\lambda_{max}\). In this case, \(\lambda_{max}=4.0359\). After this, CI is calculated using the formula: 
\begin{equation}\label{eq: 3}
    CI=\frac{\lambda_{max}-n}{n-1}=\frac{4.0359-4}{4-1}=0.120
\end{equation}
Where, \(n=4\), as 4 alternatives are considered. 
The consistency index of the matrix is compared to the consistency index of a random matrix in the final stage of AHP's calculation of the consistency ratio. A matrix with randomly input judgements is referred to as a random matrix. Consequently, it is anticipated to be very inconsistent. A calculated Random-like matrix value for matrixes of different sizes is also provided (Saaty, 1987). 
\begin{equation}\label{eq. 4}
\text{Consistency Ratio}~(CR)=\frac{CI}{\text{Random Index}}=\frac{0.120}{0.89}=0.0134<0.10
\end{equation}
Since the \(CR<0.10\), as calculated in \ref{eq. 4} we get that out input is consistent. If the comparisons are acceptable, ranking of the alternatives is done from the best to the worst based on approximate alternative score, i.e., normalized weights of each influencing factor calculated in decreasing order, also called the Criteria weight.
Factors are ranked by collecting the survey data and calculated outcomes.

\begin{table}[H]
    \centering
\renewcommand{\arraystretch}{1.2}
\setlength\tabcolsep{36pt}
    \begin{tabular}{l|cc}
    \hline
                        & \textbf{Criteria   Weight} & \textbf{Rank} \\ \hline
\textbf{Friends}        & 0.1064                     & 4             \\
\textbf{Family/Partner} & 0.3144                     & 1             \\
\textbf{Phone}          & 0.2701                     & 3             \\
\textbf{Job/Studies}    & 0.3091                     & 2            
\end{tabular}
    \caption{Ranking of AHP 1}
    \label{8}
\end{table}

According to table 4, unlike the pre-pandemic time, most of the time was devoted to the near and dear ones, followed by their jobs, mobile phones, and friends in decreasing order. People understood the importance of having a family in times of crisis. With the lockdown and work-from-home policy, people spent more time with their families than before. The pandemic limited in-person socializing and meet-ups, leading to a decrease in time spent with friends. Remote work became more prevalent, which led to the blurring of the line between work and personal time. 
With the shift from restricted eight-hour standard working hours to unrestricted timings, people were forced to give time to their jobs. The usage of phones to reduce boredom, social interactions, news updates, information, and remote work/schooling were done. 
People kept themselves busy by spending their spare time on social media and OTT applications. The impact of all these factors were greatly depending on the individual and their specific circumstances.

\subsection{Priority}
Apart from the time left from household activities and work, one might wish to devote it either for some productive learning or entertainment, as per choice. In a time when everyone thought that the world was at a break, it wasn't. Families were getting closer, spending more time getting to know and connecting emotionally with each other [21],[34]. People were searching for long-lost friends on social media platforms such as Facebook. Helping them in their times of need and sharing their thoughts. In small localities, people made friends with their neighbours. They became a part of society and supported each other through thick and thin.
From teleworking, balancing work and life, and dealing with technological challenges, to online communication with colleagues and clients individuals went through a lot during the lockdown. Although the offices and the companies were shut down due to the onset of the pandemic and the resulting Lockdown, they were doing wonders. Their employees were working day and night for the same salary from their homes which Increased the economy. As the work-from-home policy was active, colleagues came closer and became official friends.

\begin{table}[H]
\centering
\renewcommand{\arraystretch}{1.2}
\setlength\tabcolsep{15pt}
\begin{tabular}{l|cccc}
\hline
\textbf{Criteria}   & \textbf{Friends} & \textbf{Family} & \textbf{Neighbours} & \textbf{Colleagues} \\ \hline
\textbf{Friends}    & 1.0000           & 0.3198          & 1.5684              & 1.6664              \\
\textbf{Family}     & 3.1250           & 1.0000          & 2.3334              & 2.2856              \\
\textbf{Neighbours} & 0.6667           & 0.4286          & 1.0000              & 0.7822              \\
\textbf{Colleagues} & 0.6000           & 0.4286          & 1.2857              & 1.0000             
\end{tabular}%
\caption{Priority}
\label{9}
\end{table}

From the calculations, we get that \(CI=0.0287\) and thus, \(CR=0.0323<0.1\), which implies that the matrix considered is consistent.

\begin{table}[H]
    \centering
    \renewcommand{\arraystretch}{1.2}
\setlength\tabcolsep{40pt}
    \begin{tabular}{l|cc}
    \hline
                    & \textbf{Criteria   Weight} & \textbf{Rank} \\ \hline
\textbf{Friends}    & 0.2193                     & 2             \\
\textbf{Family}     & 0.4541                     & 1             \\
\textbf{Neighbours} & 0.1533                     & 4             \\
\textbf{Colleagues} & 0.1733                     & 3            
\end{tabular}
    \caption{Ranking of AHP 2}
    \label{10}
\end{table}
Based on the participants' views, the study's evaluation Criteria were enough to analyse the impact of the lockdown and how individuals spent their time with those around them. Thinking and relationships changed. Prior to the lockdown, many individuals worked long hours with little time for family, but during the lockdown, they used the opportunity to spend as much time as possible with their loved ones. Bored of seeing the same people every day, people reconnected with old friends through social media and video calls. 
Additionally, the workload was higher and the time spent in the work place was also increased, resulting in more interactions with their colleagues and co-workers. According to the results of the AHP matrix, the least time an individual spent was with their neighbours. Because of the restrictions of meeting with others, and maintaining social distancing norms, people tend to avoid meeting, and thus interacting with their neighbours. Although, they seemed to help and care for each other more during the pandemic than ever before.

\subsection{Health as a factor}
The COVID-19 pandemic has had far-reaching impacts on people's physical and mental health. Physical illness, which led to severe illness and hospitalization; mental health which led to increased stress, anxiety due to job losses, social isolation and financial stress etc.; long term effects which are yet to be discovered, but some studies suggest that there maybe some ill effects in the long-term including the possibilities of weight gain [12], respiratory and cardiovascular problems. People experienced frustration, anxiety, fear, and stress. 
During the lockdown, individuals who were confined to their homes without a proper routine and faced irregular work hours faced frustration and a lack of motivation. The repetition of being confined to one's home also impacted sleep cycles and led to unexplained laziness and fatigue [20]. The repetitive and locked life cycle of one confined to its own home also resulted in extreme troubles in the sleep cycle. The shutting of gyms, other fitness places including open gyms, morning walk parks, sports stadiums, etc., and the heightened psychological health issues have resulted in the lack of fitness of people in the lockdown. It disturbed the daily routine of exercising. 

\begin{table}[H]
    \centering
    \renewcommand{\arraystretch}{1.2}
\setlength\tabcolsep{37pt}
    \begin{tabular}{l|ccc}
    \hline
\textbf{Criteria} & \textbf{Yoga} & \textbf{Gym} & \textbf{Walk} \\
\hline
\textbf{Yoga}     & 1.00          & 1.00         & 0.50 \\
\textbf{Gym}      & 1.00          & 1.00         & 0.40 \\
\textbf{Walk}     & 2.00          & 2.50         & 1.00
\end{tabular}
    \caption{Fitness routine}
    \label{11}
\end{table}

From calculations, \(CI\) comes out to be \(0.002768\) and resulting \(CR=0.0048<0.1\), we get that AHP 3 is consistent.
\begin{table}[H]
    \centering
    \renewcommand{\arraystretch}{1.2}
    \setlength\tabcolsep{45pt}
    \begin{tabular}{l|cc}
    \hline
\textbf{}     & \textbf{Criteria   Weight} & \textbf{Rank} \\ \hline
\textbf{Yoga} & 0.24                       & 2             \\
\textbf{Gym}  & 0.23                       & 3             \\
\textbf{Walk} & 0.53                       & 1            
\end{tabular}
    \caption{Ranking of AHP 3}
    \label{12}
\end{table}

The study results were intriguing as the gym-goers who preferred gyms over any other means of physical activity were forced to do yoga or walk as they had high intensity to workout. People preferred walking, followed by yoga and gyms in decreasing order. People started considering their regular pattern and attempting to discover ways to substitute their regular activities. They started attempting to relocate their training from the gym to more convenient locations. Some opted for workouts without exercising equipment at home for maintaining their fitness routine like heavy bucket lifting, jumping jacks and skipping which require less or zero weights as an alternative to gym equipment and their availability. 
Due to the lockdown, the pollution levels also decreased as a result of which people experienced inhalation of fresh air during the day times specially early in the morning. Meditation and yoga helped in both relaxation of body and mind. The lack of motivation which the individuals experienced in the absence of gym partners started taking online yoga courses together and kept motivating each other. According to the American College of Sports Medicine, people should aim to do 150-300 minutes of aerobic exercise per week and 2 moderate-intensity strength training sessions to stay active during lockdown [20].

\subsection{Preference in time pass}
The COVID-19 pandemic had a significant impact on people's lives and resulted in many changes, both good and bad. With the restrictions of staying at home and social distancing, individuals had more time to pursue hobbies and interests that they previously couldn't fit into their busy schedules. Students took advantage of the online classes and enrolled courses which could help them in future times. More research was done in every field as far as the education sector is concerned. Some learned cook food and know they surroundings. Some individuals learned to cook, familiarized themselves with their surroundings, and took up gardening. To cope with the stress and depression caused by the news of the pandemic, people turned to their hobbies such as art, reading, and baking [24],[26]. 
As per the reports from various journals, increase in the use of social media and OTT was seen during the pandemic [6]. It became an integrated part of everyone’s life. It had both positive and negative impacts on mental and physical health.  A shift in the type of content shared on social media, with more focused topics related to pandemic such as health, news update and personal experiences. The increased usage also resulted in misinformation about the virus and its side effects. 
This analysis was important to know how people preferred to kill their time in their free time during the lockdown. 

\begin{table}[H]
    \centering
    \renewcommand{\arraystretch}{1.2}
\setlength\tabcolsep{8pt}
    \begin{tabular}{l|cccc}
    \hline
\textbf{Criteria}        & \textbf{OTT} & \textbf{Social media} & \textbf{Hobbies} & \textbf{Extra knowledge} \\ \hline
\textbf{OTT}             & 1.0000       & 0.8334                & 0.6667           & 0.6667                   \\
\textbf{Social media}    & 1.2000       & 1.0000                & 0.7778           & 0.7143                   \\
\textbf{Hobbies}         & 1.5000       & 1.2857                & 1.0000           & 0.8889                   \\
\textbf{Extra knowledge} & 1.5000       & 1.4000                & 1.1250           & 1.0000                  
\end{tabular}
    \caption{Preference in time pass}
    \label{13}
\end{table}

For the comparison matrix 4, \(CI=0.0006\) and \(CR=0.0007<0.1\), we get that AHP 4 is consistent.
\begin{table}[H]
    \centering
    \renewcommand{\arraystretch}{1.2}
\setlength\tabcolsep{35pt}
    \begin{tabular}{l|cc}
    \hline
\textbf{}                & \textbf{Criteria   Weight} & \textbf{Rank} \\ \hline
\textbf{OTT}             & 0.1918                     & 4             \\
\textbf{Social media}    & 0.2221                     & 3             \\
\textbf{Hobbies}         & 0.2813                     & 2             \\
\textbf{Extra knowledge} & 0.3048                     & 1            
\end{tabular}
    \caption{Ranking of AHP 4}
    \label{14}
\end{table}

As most of our respondents were students and those related to the field of education and research, it was not surprising that they utilized the lockdown to enhance their knowledge and skills. A significant number of respondents reported spending their time exploring new subjects, learning new skills such as coding skills, cooking, cake baking etc. Next to it, people claimed that they decided to give wings to their passion or hobbies which was dormant because of the hectic and busy schedule of daily lives. 
Regarding social media, the usage increased during the pandemic. However, research shows an increase in the cases of cyber bullying, providing invalid or incorrect information especially during the pandemic which caused mental stress and depression, even in the youth, resulting in avoiding the social media. The least time was spent in watching TV or OTT platforms such as Netflix, amazon prime video etc. Although the purchase of OTT platforms spiked at the start of the lockdown due to the availability of cheap smartphones and internet data, users soon grew tired of the same content and resulted in a decrease in usage. During the lockdown, people were tired of watching the same content in different forms and thus it resulted in decrease in the usage of OTT platforms. The televisions were merely a means of important information but excessive viewing has been linked to depression and stress.

\chapter{CONCLUSIONS}
The COVID-19 pandemic has had a major impact on the world, including India, and its effects on the economy and human lives are still being felt. Although many studies have been conducted on the topic, there is a lack of research on the impact of the pandemic on human behaviour. This study aims to address this gap by examining the changes in human behaviour resulting from the COVID-19 pandemic. The methodology of the study involved an online survey, and the results showed that people's attitudes towards their family and friends changed, as they devoted more time to spending with loved ones. People also became more aware of the importance of life insurance and planning for the future after facing economic difficulties and the loss of loved ones. Working from home and online teaching changed people's perspectives on their jobs and education, and the lockdown resulted in a change in people's sleeping patterns, with majority respondents sleeping more. Boredom and a lack of entertainment led to an increase in food intake, but people opted for healthy home-made food over junk food or skipping the food. The lockdown also intensified the effects on people's mental health, causing increased stress and anxiety due to fear of infection. However, people tried to maintain good health and immunity through activities like yoga and home workouts. Despite some limitations in the self-reported data, this study provides valuable insights into the impact of the COVID-19 lockdown on human behaviour. Future research should focus on a more detailed examination of the effects of COVID-19 on a larger sample of the population.

\section{Future Scopes}
The COVID-19's death toll has been rising over time. Despite stringent safety precautions like the lockdown of the majority of the afflicted cities, movement restrictions, the use of appropriate masks, and so on, it has crossed all boundaries and borders. However, dozens of new instances are reported every day. The inability to control the deadly COVID-19 outbreak has caused problems for policymakers and medical personnel.
Emphasizing the complexity of human behaviour and health-related issues is crucial. Most empirical research relies on standard statistical methods which are limited in their ability to understand human behaviour. The use of big data techniques may help identify the most critical characteristics due to the complexity of human behaviour theories and the presence of various factors that influence human conduct. Clustering is the process of dividing a group of people or objects into smaller groups. A cluster is a group of data that is distinct from one another yet related. The more data that is collected, the greater the understanding of human behaviour will be [25]. The availability of big data technology and the decreasing cost of data collection by businesses make it possible to create a comprehensive picture of humanity at a subject level. The authors suggest that there is strong justification for further research in this field. It is crucial to adopt a fresh and comprehensive approach to understand the interconnection between genetics, the mind-body relationship, behaviour, and the environment with human health. According to a research article [2], big data approaches have the potential to speed up discovery and lead to the formation of new theories directly from data, rather than relying on theory-driven methodologies commonly used in psychological studies.

\section{Limitations}
Our results maybe biased due to the nature of the online survey and the respondents were largely the young respondents. 
The relationship between statistics and behaviour analysis has been complex and challenging. While there has been a gradual shift from visual inspection to null hypothesis testing and quantitative modelling, aggregating dependent variables can reduce the accuracy of statistical analysis and hinder the proper depiction of existing variability. This highlights the limitations of human perception and judgment in behaviour analysis. Hence, there is a need for contemporary statistical analysis, including the use of model comparison, multilevel modelling, and generalized linear modelling, in conjunction with stronger statistical training in behaviour analysis programs.
Behaviour analysis has traditionally relied on experimental methods and visual inspection to control extraneous sources of variation and assess the effects of interventions. However, the increasing use of quantitative methods in the study of behaviour has led to a greater use of statistical tests in behaviour analysis.

\begin{appendices}
\chapter{Survey Form}

\begin{figure}[H]
    \centering
    \includegraphics[width=\textwidth, keepaspectratio]{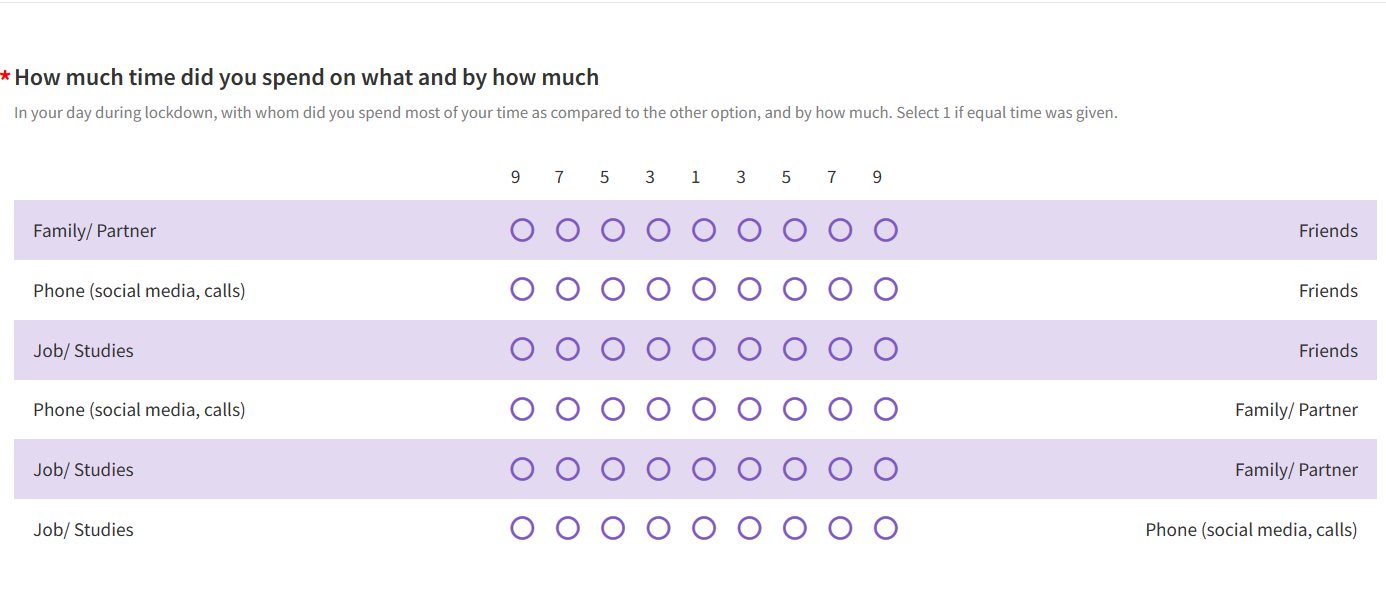}
   
\end{figure}

\begin{figure}[H]
    \centering
    \includegraphics[width=\textwidth, keepaspectratio]{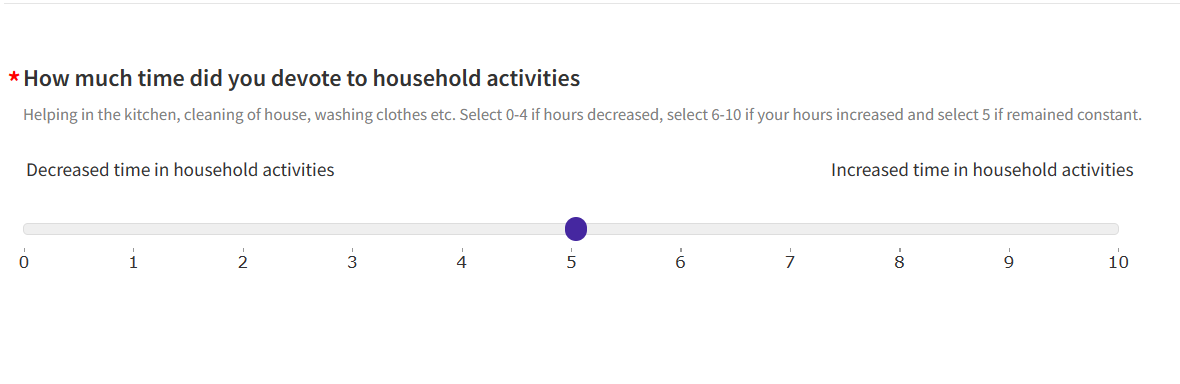}

\end{figure}

\begin{figure}[H]
    \centering
    \includegraphics[width=\textwidth, keepaspectratio]{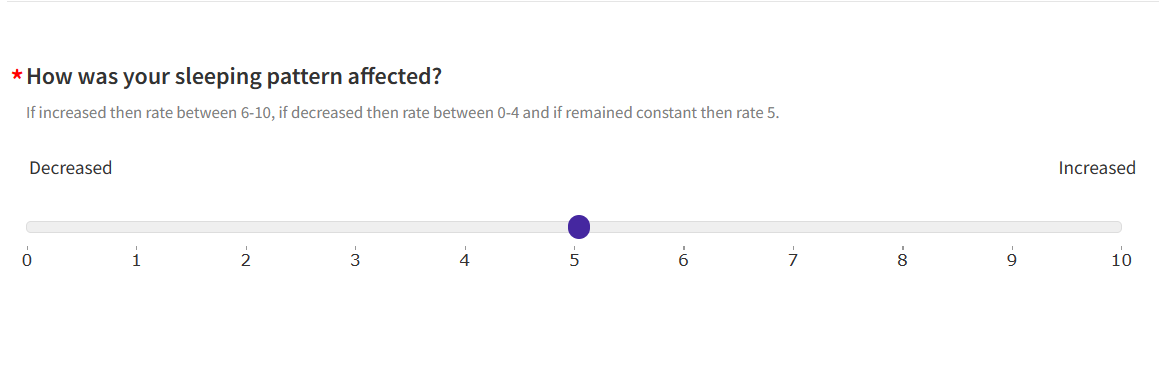}
    
\end{figure}

\begin{figure}[H]
    \centering
    \includegraphics[width=\textwidth, keepaspectratio]{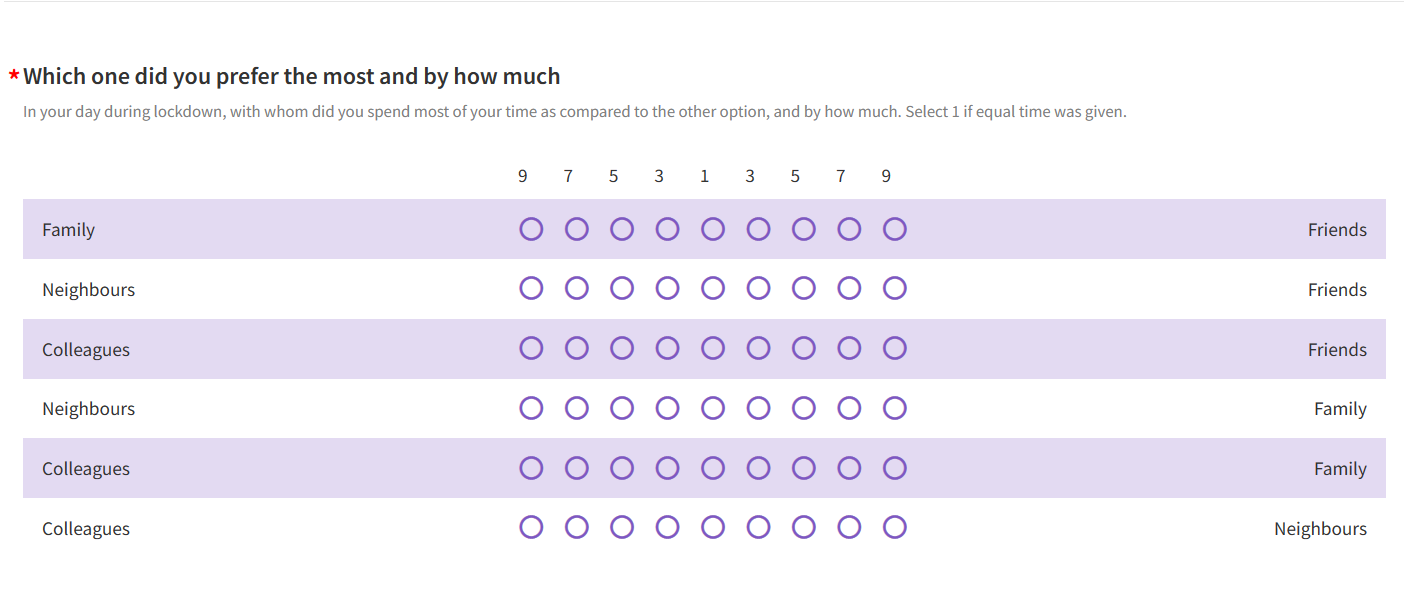}
    
\end{figure}

\begin{figure}[H]
    \centering
    \includegraphics[width=\textwidth, keepaspectratio]{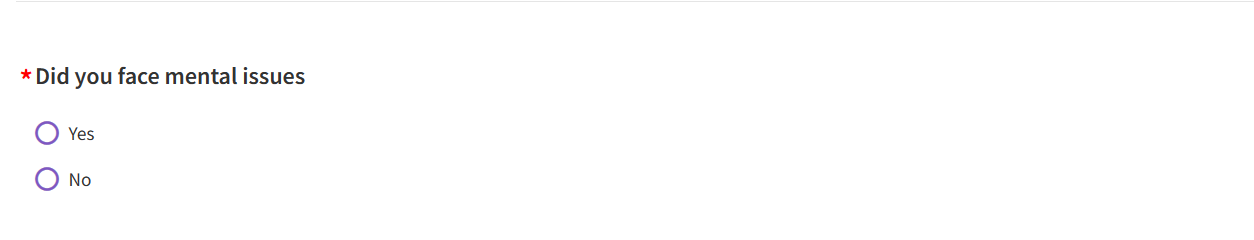}

\end{figure}

\begin{figure}[H]
    \centering
    \includegraphics[width=\textwidth, keepaspectratio]{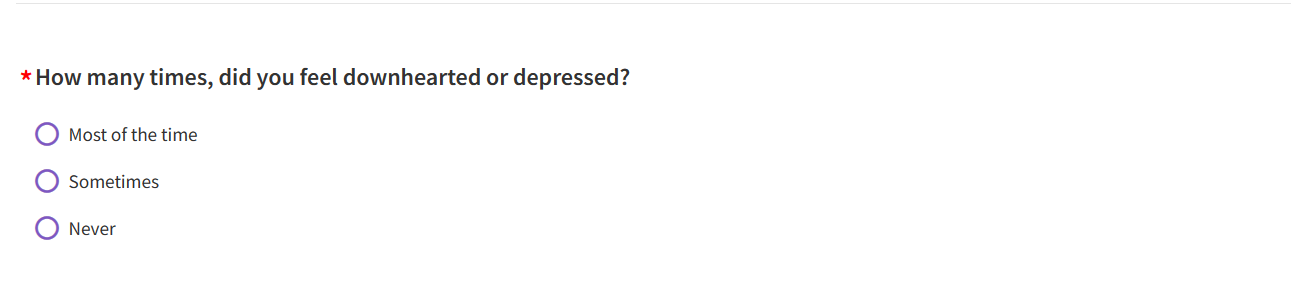}

\end{figure}

\begin{figure}[H]
    \centering
    \includegraphics[width=\textwidth, keepaspectratio]{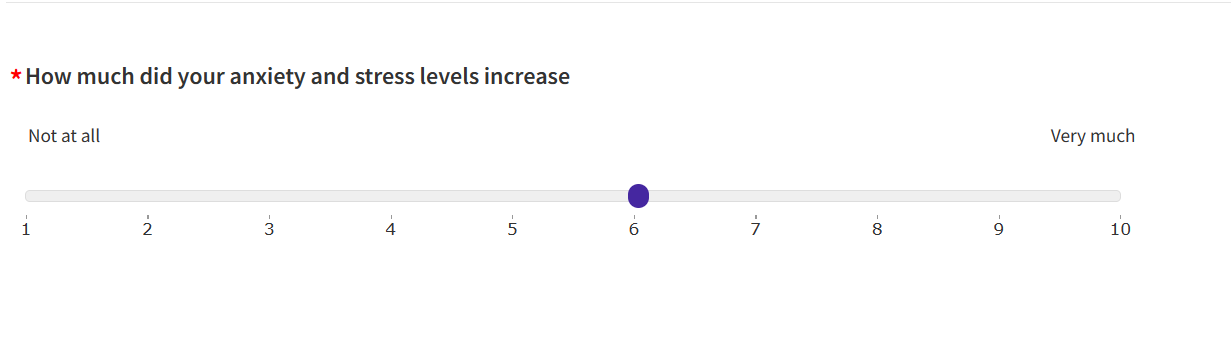}

\end{figure}

\begin{figure}[H]
    \centering
    \includegraphics[width=\textwidth, keepaspectratio]{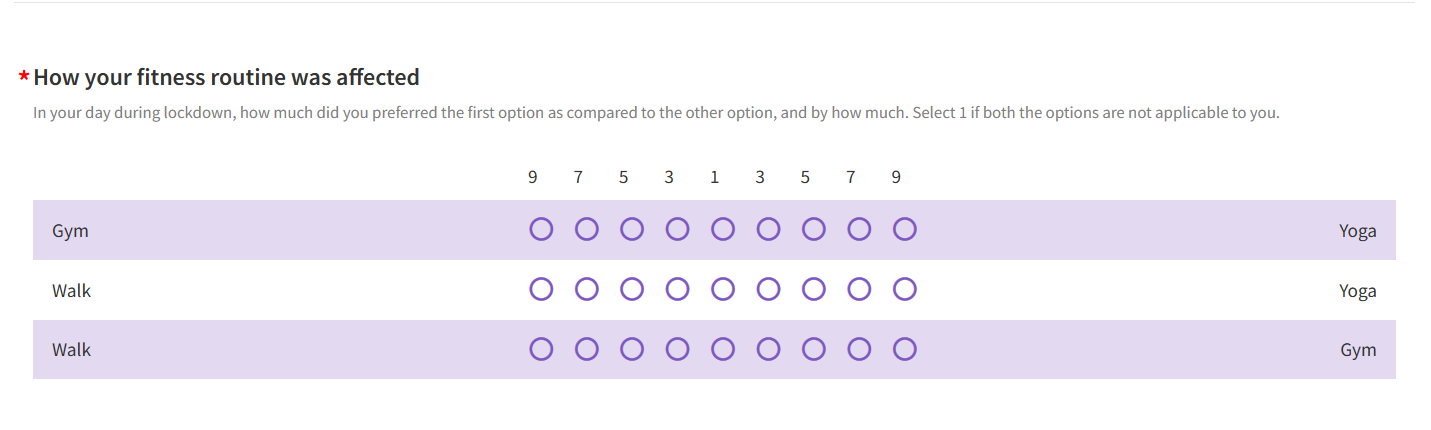}

\end{figure}

\begin{figure}[H]
    \centering
    \includegraphics[width=\textwidth, keepaspectratio]{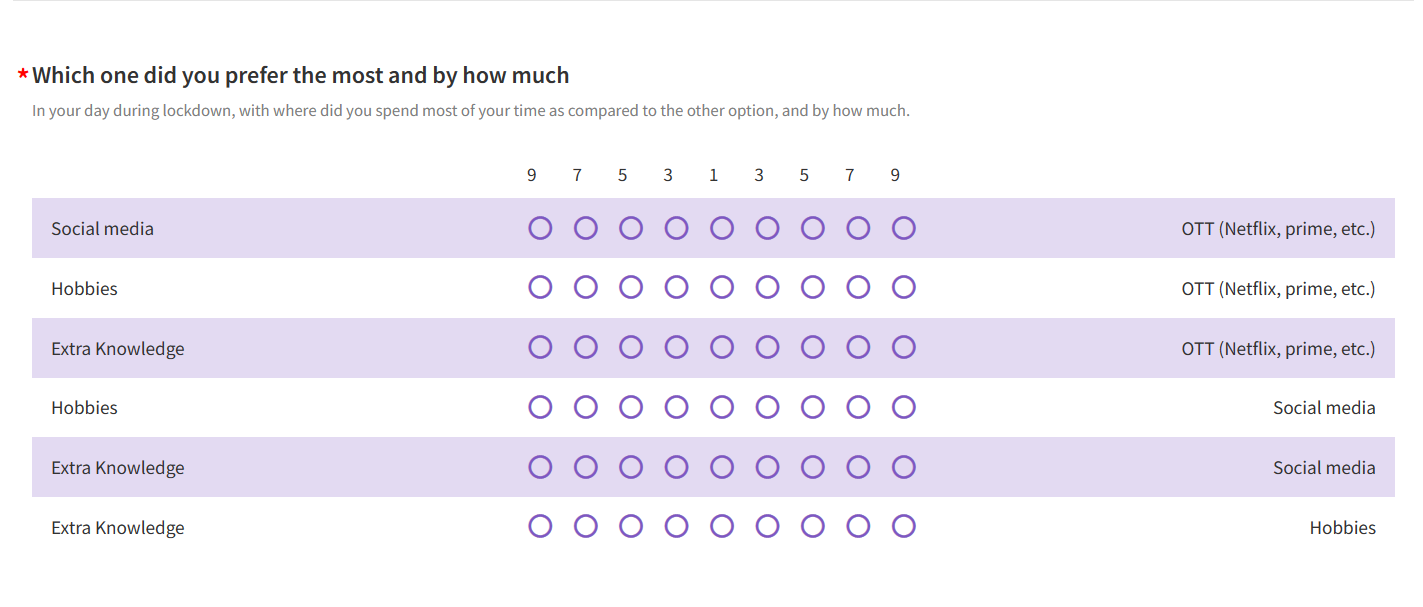}

\end{figure}

\begin{figure}[H]
    \centering
    \includegraphics[width=\textwidth, keepaspectratio]{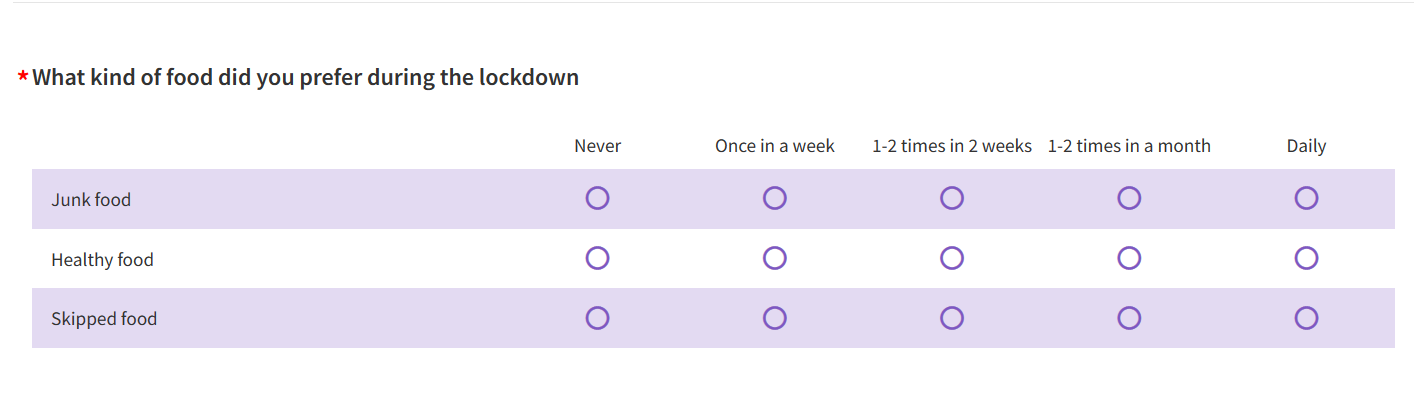}

\end{figure}

\chapter{Excel formulations}
Excel formulation and calculations:
\begin{enumerate}
    \item  Insert the data in the upper triangular matrix row-wise.
    \item Sum all the columns using \textbf{\(=SUM(B2:B5)\) }as for column B, write in row \(6\).
    \item Calculate \textbf{Priority vector (PV)} using\textbf{ \(=(B2*C2*D2*E2)^{(1/4)}\)} as in column \(F\).
    \item Calculate the sum \(F2:F5\)  using \(=SUM(F2:F5)\) in and write the result in \(F6\).
    \item Calculate \textbf{Normalized vector (NW)} for Friends using \(=F2/\$F\$6\), and similarly for other criteria. The sum of all NW should come out to be 1, as in \(G6\) cell.
    \item Calculate \textbf{Eigen vector (EV)} for Friends using \(=(B2*\$G\$2)+(C2*\$G\$3+D2*\$G\$4+E2*\$G\$5)\), similarly for other criteria. 
    \item Evaluate \textbf{Consistency Measure (CM)} using \(=H2/G2\) which is NW/EW from each criteria respectively.
    \item Calculate the average of CM using \(=AVERAGE(I2:I5)\) which is the value of \textbf{Lam\_max}.
    \item Evaluate \textbf{Consistency index (CI)} using \(=(I8-4)/(4-1)\) where I8 is \textbf{Lam\_max}.
    \item Calculate \textbf{Consistency ratio (CR)} using \(=E9/0.89\) as \(n=4\).
    \item Since CR \(<0.1\), we say that the AHP is consistent. 
\end{enumerate}

\begin{figure}[H]
    \centering
    \includegraphics[width=\textwidth, keepaspectratio]{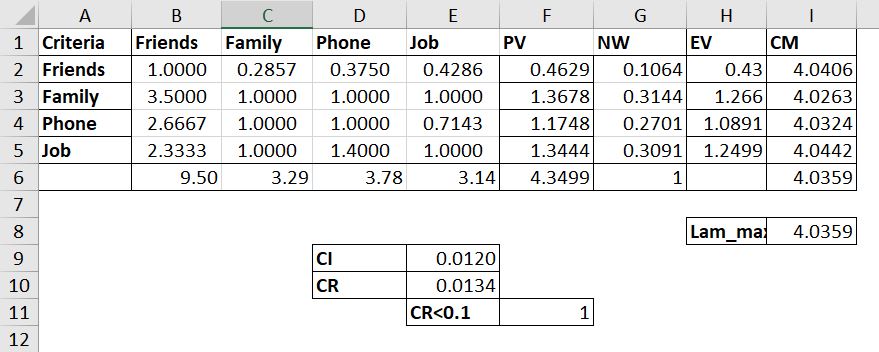}
    \caption{Excel Output}
    \label{Excel Output}
\end{figure}
\end{appendices}


\begin{thebibliography}{99}

\bibitem{1}
Ahmad, T., Haroon, Baig, M., Hui, J., 2020. Coronavirus disease 2019 (Covid-19) pandemic and economic impact. Pakistan J. Med. Sci. 36, S73–S78. https://doi.org/10.12669/pjms.36.COVID19-S4.2638

\bibitem{2}
Azmak, O., Bayer, H., Caplin, A., Chun, M., Glimcher, P., Koonin, S., Patrinos, A., 2015. Using big data to understand the human condition: The Kavli HUMAN project. Big Data 3, 173–188. https://doi.org/10.1089/big.2015.0012

\bibitem{3}
Bavel, J.J.V., Baicker, K., Boggio, P.S., Capraro, V., Cichocka, A., Cikara, M., Crockett, M.J., Crum, A.J., Douglas, K.M., Druckman, J.N., Drury, J., Dube, O., Ellemers, N., Finkel, E.J., Fowler, J.H., Gelfand, M., Han, S., Haslam, S.A., Jetten, J., Kitayama, S., Mobbs, D., Napper, L.E., Packer, D.J., Pennycook, G., Peters, E., Petty, R.E., Rand, D.G., Reicher, S.D., Schnall, S., Shariff, A., Skitka, L.J., Smith, S.S., Sunstein, C.R., Tabri, N., Tucker, J.A., Linden, S. van der, Lange, P. van, Weeden, K.A., Wohl, M.J.A., Zaki, J., Zion, S.R., Willer, R., 2020. Using social and behavioural science to support COVID-19 pandemic response. Nat. Hum. Behav. 4, 460–471. https://doi.org/10.1038/s41562-020-0884-z

\bibitem{4}
Biroli, P., Bosworth, S., Della Giusta, M., Di Girolamo, A., Jaworska, S., Vollen, J., 2021. Family Life in Lockdown. Front. Psychol. 12, 1–13. https://doi.org/10.3389/fpsyg.2021.687570
Ahmad, T., Haroon, Baig, M., Hui, J., 2020. Coronavirus disease 2019 (Covid-19) pandemic and economic impact. Pakistan J. Med. Sci. 36, S73–S78. https://doi.org/10.12669/pjms.36.COVID19-S4.2638

\bibitem{5}
Bruni, O., Malorgio, E., Doria, M., Finotti, E., Spruyt, K., Melegari, M.G., Villa, M.P., Ferri, R., 2022. Changes in sleep patterns and disturbances in children and adolescents in Italy during the Covid-19 outbreak. Sleep Med. 91, 166–174. https://doi.org/10.1016/j.sleep.2021.02.003

\bibitem{6}
Chebrolu, R.H., Janagam, J., Muraleedharan, K.C., R., R., 2021. Impact of social media and over the top media during COVID-19 lockdown, a cross-sectional study. Int. J. Community Med. Public Heal. 8, 1156. https://doi.org/10.18203/2394-6040.ijcmph20210796

\bibitem{7}
Cho, Y.Y., Woo, H., 2022. Factors in Evaluating Online Learning in Higher Education in the Era of a New Normal Derived from an Analytic Hierarchy Process (AHP) Based Survey in South Korea. Sustain. 14. https://doi.org/10.3390/su14053066

\bibitem{8}
Christakis, N.A., Fowler, J.H., 2013. Social contagion theory: Examining dynamic social networks and humanbehavior. Stat. Med. 32, 556–577. https://doi.org/10.1002/sim.5408

\bibitem{9}
Compagno, L., D’Urso, D., Latora, A.G., Trapani, N., 2012. Influence of AHP methodology and human behaviour on e-scouting process. IFIP Adv. Inf. Commun. Technol. 384 AICT, 514–525. https://doi.org/10.1007/978-3-642-33980-6-56

\bibitem{10}
Das, D., Datta, A., Kumar, P., Kazancoglu, Y., Ram, M., 2022. Building supply chain resilience in the era of COVID-19: An AHP-DEMATEL approach. Oper. Manag. Res. 15, 249–267. https://doi.org/10.1007/s12063-021-00200-4

\bibitem{11}
Dodd, R.H., Dadaczynski, K., Okan, O., McCaffery, K.J., Pickles, K., 2021. Psychological wellbeing and academic experience of university students in australia during covid-19. Int. J. Environ. Res. Public Health 18, 1–12. https://doi.org/10.3390/ijerph18030866

\bibitem{12}
Dor-Haim, H., Katzburg, S., Revach, P., Levine, H., Barak, S., 2021. The impact of COVID-19 lockdown on physical activity and weight gain among active adult population in Israel: a cross-sectional study. BMC Public Health 21, 1–10. https://doi.org/10.1186/s12889-021-11523-z

\bibitem{13}
Eftimov, T., Popovski, G., Petković, M., Seljak, B.K., Kocev, D., 2020. COVID-19 pandemic changes the food consumption patterns. Trends Food Sci. Technol. 104, 268–272. https://doi.org/10.1016/j.tifs.2020.08.017

\bibitem{14}
Francas, D., Minner, S., 2009. Manufacturing network configuration in supply chains with product recovery. Omega 37, 757–769. https://doi.org/10.1016/j.omega.2008.07.007

\bibitem{15}
Gorbalenya, A.E., Baker, S.C., Baric, R.S., Groot, R.J. De, Gulyaeva, A.A., Haagmans, B.L., Lauber, C., Leontovich, A.M., 2020. The species and its viruses – a statement of the oronavirus study group. Biorxiv (Cold Spring Harb. Lab. 1–15.

\bibitem{16}
Howard, M., Hopkinson, P., Miemczyk, J., 2019. The regenerative supply chain: a framework for developing circular economy indicators. Int. J. Prod. Res. 57, 7300–7318. https://doi.org/10.1080/00207543.2018.1524166

\bibitem{17}
Hsiao, S.W., 2002. Concurrent design method for developing a new product. Int. J. Ind. Ergon. 29, 41–55. https://doi.org/10.1016/S0169-8141(01)00048-8

\bibitem{18}
Hussain, M.W., Mirza, T., Hassan, M.M., 2020. Impact of COVID-19 Pandemic on the Human Behavior. Int. J. Educ. Manag. Eng. 10, 35–61. https://doi.org/10.5815/ijeme.2020.05.05

\bibitem{19}

Kandeger, A., Guler, H.A., Egilmez, U., Guler, O., 2018. Major depressive disorder comorbid severe hydrocephalus caused by Arnold – Chiari malformation Does exposure to a seclusion and restraint event during clerkship influence medical student ’ s attitudes toward psychiatry  Indian J. Psychiatry 59, 2017–2018. https://doi.org/10.4103/psychiatry.IndianJPsychiatry

\bibitem{20}
Kaur, H., Singh, T., Arya, Y.K., Mittal, S., 2020. Physical Fitness and Exercise During the COVID-19 Pandemic: A Qualitative Enquiry. Front. Psychol. 11, 1–10. https://doi.org/10.3389/fpsyg.2020.590172

\bibitem{21}
Kutsar, D., Kurvet-Käosaar, L., 2021. The Impact of the COVID-19 Pandemic on Families: Young People’s Experiences in Estonia. Front. Sociol. 6, 1–12. https://doi.org/10.3389/fsoc.2021.732984

\bibitem{22}
Ministry of Health \& Family Welfare, 2020. MoHFW Home. Minist. Heal. Fam. Welfare, Gov. India.

\bibitem{23}
Montesi, M., 2021. Human information behavior during the Covid-19 health crisis. A literature review. Libr. Inf. Sci. Res. 43, 101122. https://doi.org/10.1016/j.lisr.2021.101122

\bibitem{24}
Morse, K.F., Fine, P.A., Friedlander, K.J., 2021. Creativity and Leisure During COVID-19: Examining the Relationship Between Leisure Activities, Motivations, and Psychological Well-Being. Front. Psychol. 12. https://doi.org/10.3389/fpsyg.2021.609967

\bibitem{25}
Moustafa, A.A., Diallo, T.M.O., Amoroso, N., Zaki, N., 2018. Applying Big Data Methods to Understanding Human Behavior and Health 12, 10–13. https://doi.org/10.3389/fncom.2018.00084

\bibitem{26}
Panarese, P., Azzarita, V., 2021. The Impact of the COVID-19 Pandemic on Lifestyle: How Young people have Adapted Their Leisure and Routine during Lockdown in Italy. Young 29, S35–S64. https://doi.org/10.1177/11033088211031389
\bibitem{27}
Pfefferbaum, B., North, C.S., 2020. Mental Health and the Covid-19 Pandemic. N. Engl. J. Med. 383, 510–512. https://doi.org/10.1056/nejmp2008017

\bibitem{28}
Rehman, U., Shahnawaz, M.G., Khan, N.H., Kharshiing, K.D., Khursheed, M., Gupta, K., Kashyap, D., Uniyal, R., 2021. Depression, Anxiety and Stress Among Indians in Times of Covid-19 Lockdown. Community Ment. Health J. 57, 42–48. https://doi.org/10.1007/s10597-020-00664-x

\bibitem{29}
Saaty, R.W., 1987. The analytic hierarchy process-what it is and how it is used. Math. Model. 9, 161–176. https://doi.org/10.1016/0270-0255(87)90473-8

\bibitem{30}
Shah, A.K., Ravichandran, Prabhadevi, Ravichandran, Prabhu, 2020. COVID-19 pandemic: insights into human behaviour. Int. J. Community Med. Public Heal. 7, 4213. https://doi.org/10.18203/2394-6040.ijcmph20204399

\bibitem{31}
Slama, Hela, El Kefi, H., Taamallah, K., Stambouli, N., Baffoun, A., Samoud, W., Bechikh, C., Oumaya, A., Lamine, K., Hmida, M.J., Slama, Hichem, Ferjani, M., Gharsallah, H., 2021. Immediate Psychological Responses, Stress Factors, and Coping Behaviors in Military Health-Care Professionals During the COVID-19 Pandemic in Tunisia. Front. Psychiatry 12, 1–11. https://doi.org/10.3389/fpsyt.2021.622830

\bibitem{32}
Stamu-O’Brien, C., Carniciu, S., Halvorsen, E., Jafferany, M., 2020. Psychological aspects of COVID-19. J. Cosmet. Dermatol. 19, 2169–2173. https://doi.org/10.1111/jocd.13601

\bibitem{33}
Taherdoost, H., 2017. Decision Making Using the Analytic Hierarchy Process ( AHP ); A Step by Step Approach. J. Econ. Manag. Syst. 2, 244–246.

\bibitem{34}
Vanderhout, S.M., Birken, C.S., Wong, P., Kelleher, S., Weir, S., Maguire, J.L., 2020. Family perspectives of COVID-19 research. Res. Involv. Engagem. 6, 4–6. https://doi.org/10.1186/s40900-020-00242-1

\bibitem{35}
WHO, 2020. Coronavirus disease 2019 (2019-nCOV) Situation Report – 11. World Heal. Organ. 1–7.

\bibitem{36}
Wiederhold, B.K., 2020. Children’s Screen Time during the COVID-19 Pandemic: Boundaries and Etiquette. Cyberpsychology, Behav. Soc. Netw. 23, 359–360. https://doi.org/10.1089/cyber.2020.29185.bkw

\bibitem{37}
Wind, Y., Saaty, T.L., 1980. 8002 Marketing Applications of the Analytic.pdf  Manage. Sci.

\bibitem{38}
Zhang, Y., 2020. Spending Time with Family Members: How COVID-19 has Changed the Family Member Relationship 496, 173–182. https://doi.org/10.2991/assehr.k.201214.488

\bibitem{39}
Zuraidi, S.N.F., Rahman, M.A.A., Akasah, Z.A., 2018. A Study of using AHP Method to Evaluate the Criteria and Attribute of Defects in Heritage Building. E3S Web Conf. 65, 1–14. https://doi.org/10.1051/e3sconf/20186501002


\end{thebibliography}
\end{document}